%% file: arxiv.tex
\documentclass[letterpaper,twocolumn,10pt]{article}
\usepackage{usenix2019_v3}
\usepackage{epsfig,endnotes}
\usepackage{enumitem}
\usepackage{balance}
\usepackage{tabularx}
\usepackage{float}

\usepackage{breakurl}
\usepackage{color}
\usepackage{booktabs}
\usepackage{adjustbox}
\usepackage{multirow}
\usepackage{multicol}
\usepackage{makecell}
\usepackage{subcaption}
\usepackage{verbatim}
\usepackage{comment}
\usepackage{caption}
\usepackage{xpatch}
\usepackage{colortbl}
\usepackage{amsmath}
\usepackage{xspace}
\usepackage[normalem]{ulem}

\useunder{\uline}{\ul}{}
\definecolor{Orange}{rgb}{1,0.5,0}
\definecolor{airforceblue}{rgb}{0.36, 0.54, 0.66}
\definecolor{amethyst}{rgb}{0.6, 0.4, 0.8}

\newcommand{\citep}[1]{\cite{#1}}

\newcommand{\parb}[1]{\vspace{0.05in}\noindent{\textbf{#1}}\quad}
\newcommand{\parc}[1]{\vspace{0.05in}\noindent{\textbf{#1}}\;}

\newcommand{\clickbait}{\textsf{Clickbait}\xspace{}}
\newcommand{\prohibited}{\textsf{Potentially Prohibited}\xspace{}}
\newcommand{\deceptive}{\textsf{Deceptive}\xspace{}}
\newcommand{\sensitive}{\textsf{Sensitive}\xspace{}}
\newcommand{\sensitiveother}{\textsf{Sensitive: Other}\xspace{}}
\newcommand{\sensitivefinancial}{\textsf{Sensitive: Financial}\xspace{}}
\newcommand{\neutral}{\textsf{Neutral}\xspace{}}
\newcommand{\opportunity}{\textsf{Opportunity}\xspace{}}
\newcommand{\healthcare}{\textsf{Healthcare}\xspace{}}
\newcommand{\political}{\textsf{Political}\xspace{}}
\newcommand{\problematic}{\textsf{Problematic}\xspace{}}

\newenvironment{packed_itemize}{
\begin{itemize}
  \setlength{\itemsep}{.5pt}
  \setlength{\parskip}{0pt}
  \setlength{\parsep}{0pt}
  \setlength{\topsep}{1pt}
  \setlength{\itemindent}{0pt}
}{\end{itemize}}

\usepackage{dcolumn}
\newcolumntype{d}[1]{D{.}{.}{#1}}

\begin{document}

\pagenumbering{gobble}

%don't want date printed
\date{}

%make title bold and 14 pt font (Latex default is non-bold, 16 pt)
\title{\Large \bf Problematic Advertising and its Disparate Exposure on Facebook}

\author{
\and
\and
\and
{\rm Muhammad Ali}\\
Northeastern University\\
% mali@ccs.neu.edu
\and
{\rm Angelica Goetzen}\\
Max Planck\\
Institute for Software Systems\\
% agoetzen@mpi-sws.org 
\and \and \and \and
 {\rm Alan Mislove}\\
 Northeastern University\\
%  amislove@ccs.neu.edu
 \and
 {\rm Elissa M. Redmiles}\\
 Max Planck\\
 Institute for Software Systems\\
%  eredmiles@mpi-sws.org
 \and
 {\rm Piotr Sapiezynski}\\
 Northeastern University\\
%  p.sapiezynski@northeastern.edu
 %Name Institution
} % end author

\maketitle

\input{abstract}
\input{intro}
\input{related}
\input{methods}
\input{results}

\input{discussion}

\section*{Acknowledgements}
We are grateful to our shepherd and reviewers for their valuable feedback.
We also thank our annotators, Devesh Tarasia and Manjot Bedi, for their work.
This work is funded in part by NSF grants CNS-1916020 and CNS-1955227, and Mozilla Research Grant 2019H1.

%\balance
{\footnotesize \bibliographystyle{acm}
\bibliography{references}}

\input{arxiv_appendix}
\input{arxiv-ae}

\end{document}

%% file: abstract.tex
%!TEX root = main.tex
\begin{abstract}
% Problematic media have long captivated researchers' attention for their potential to cause negative outcomes for viewers. Given the prevalence of online targeted advertising as a media form, it is thus important to understand the problematic experiences users have with online ads.
% Recent work has explored people's sentiments toward online ads and the impacts of these ads on people's online experiences, finding evidence that online ads can indeed be problematic.
Targeted advertising remains an important part of the free web browsing experience, where advertisers' targeting and personalization algorithms together find the most relevant audience for millions of ads every day.
However, given the wide use of advertising, this also enables using ads as a vehicle for problematic content, such as scams or clickbait.
Recent work that explores people's sentiments toward online ads, and the impacts of these ads on people's online experiences, has found evidence that online ads can indeed be problematic.
Further, there is the potential for personalization to aid the delivery of such ads, even when the advertiser targets with low specificity.
In this paper, we study Facebook---one of the internet’s largest ad platforms---and investigate key gaps in our understanding of problematic online advertising: (a) What categories of ads do people find problematic? (b) Are there disparities in the distribution of problematic ads to viewers? and if so, (c) Who is responsible---advertisers or advertising platforms?
To answer these questions, we empirically measure a diverse sample of user experiences with Facebook ads via a 3-month longitudinal panel.
We categorize over 32,000 ads collected from this panel ($n=132$); and survey participants' sentiments toward their own ads to identify four categories of problematic ads.
Statistically modeling the distribution of problematic ads across demographics, we find that older people and minority groups are especially likely to be shown such ads. Further, given that
%fewer than 25\% of problematic ads were targeted to specific demographics by advertisers
22\% of problematic ads had no specific targeting from advertisers, we infer that ad delivery algorithms (advertising platforms themselves) played a significant role in the biased distribution of these ads.

\end{abstract}

%% file: intro.tex
%!TEX root = main.tex
\section{Introduction}
%\elissa{@angelica please try to condense where possible -- paragraphs 1 and 2 are prime targets; would like it no longer than the first column on page 2}
Targeted advertising fuels a sizable part of the web's economy today~\cite{evans2009online}. %, allowing many parts of the web to remain free (e.g., many news sites and social media platforms).
Behind the ads shown on digital platforms are complex marketplaces where advertisers compete for user attention, and advertising platforms such as Google, Facebook, and Twitter---capitalizing on user data---act as intermediaries.
To identify the right audience for each ad, these platforms provide detailed targeting options to advertisers, as well as sophisticated personalization algorithms designed to %supplement the targeting criteria and 
find the most ``relevant'' audience.
As a result, the ads that constitute a user's everyday experience are determined by a confluence of factors: what time the user is browsing, which advertisers were trying to target them, and what content the platform's personalization algorithm considers relevant to them.
Further, due to the scale of these marketplaces, users run into ads on a vast variety of topics---ranging from neutral product ads, to opportunity ads for jobs and scholarships, and even to problematic clickbait ads and scams.

Given the wide variance in ads a user may potentially receive, it is important to consider whether %the distribution of potentially problematic ads on platforms, and how 
some users' {\em overall ad experience} might be worse than others. % through increased exposure to problematic ads. 
Prior work has illustrated the impact of harmful media \cite{smith1998harmful,wakefield2003role,vidmar1974archie,browne1999television,van2007dieting,posavac1998exposure,aubrey2006exposure}, has theorized about the ways 
%\elissa{theoreized about? needs to contrast with the next sentence as it's currently written or next sentence needs to be rephrased} 
in which digital ads may harm users~\cite{li2012knowing,nelms2016towards,rastogi2016these,zarras2014dark,pengnate2016measuring,scott2021you,gak2022distressing}, and %There has been a number of  works~\cite{zeng2021polls,xie2011susceptible,nelms2016towards,xie2015disentangling,gak2022distressing,gomez2020fail,pengnate2021effects,fb1} that have brought attention to ads that might be problematic to users in different ways.
has asked users themselves to express why they find certain ads problematic~\cite{zeng2021makes}. %Further recent work has studied the impact of ads on sensitive populations, such as those with mental health disorders, who may experience deeper levels of harm from ads that target their sensitivities \cite{gak2022distressing}.
However, a complete understanding on the online ad experiences of individual users, along with a breakdown of the kinds of ads different users find problematic, remains elusive.

%These works provide evidence that there exist ads in the targeted advertising landscape that users dislike and find problematic, not merely for being unsolicited content, but for being untrustworthy, deceptive, or distressing. 
%

In this paper, we build on prior work to systematically identify which categories of ads people perceive as problematic, evaluate if there are skews in the delivery of problematic categories of ads, and determine the roles of advertisers and personalization algorithms in the distribution of problematic ads.
Thus, we aim to answer the following research questions: 
\begin{enumerate}[label={\bfseries RQ\arabic*:},leftmargin = 3em,itemsep = 0.5pt]
    \item What categories of ads are perceived as problematic?
    \item Are there skews in the distribution of problematic ads?
    \item Who is responsible for any observed skews?
\end{enumerate}

% \begin{packed_itemize}
%     \item What categories of ads do people perceive as problematic?
%     \item Are there skews in the distribution of problematic ads (defined in RQ1)?
%     \item Who is responsible for any skews identified in RQ2?
% \end{packed_itemize}

%we address this situation by examining how users {\em themselves} view potentially problematic ads, and investigate how such ads are distributed across the user population.
%
%For example, are these ads delivered equally to all users, or does a subset of the population have a uniquely poor ad experience?
%

To do so, we recruit a panel of $132$ paid participants, who we select across a variety of demographic categories. We longitudinally observe participants' Facebook ad experiences over a period of three months, collecting the ads they receive, and the revealed targeting information for each ad. 
We choose Facebook as our platform of study because it is one of the largest and most data-rich personalized advertising platforms. %; it also has a relatively thin line between native content and advertising, with both embedded into the same feed. \elissa{explain better why this thin line is important?}
We use a combination of (1) logged data and (2) quantitative surveys to measure our participants' ad experiences. First, we instrument our participants' web browsers to collect all Facebook ads they are shown in their desktop browsers, alongside the detailed targeting information Facebook provides for these ads. Second, using a combination of inductive qualitative coding, and deductive analysis of computational and social science research, as well as existing platform policies, we develop a {\it codebook} of ad categories, covering a variety of potentially problematic ad types. Using human raters, we then classify over 32,000 ads shown to our participants using this codebook. With this coded data, we regularly survey our participants to assess which types of ads---within the set of ads that they are shown by Facebook and which we annotated---they find problematic and why. %
%We find a number of expected and surprising trends.

Using the collected data, we first examine the content that participants %\elissa{say participants not users any place you're talking about your data} 
dislike (RQ1). We identify four categories of ads that participants find problematic (i.e., are disliked more than ads of any other category): deceptive ads, content that is prohibited by Facebook itself, clickbait, and ads considered sensitive by platform or government policy (e.g., ads for weight loss, gambling or alcohol).  %We find that 
%other ad types such as those related to educational programs and health studies, which could potentially be problematic due to advertising low quality or high cost opportunities, are actually disliked at a rate similar to ads classified as neutral.
%

% percent problematic for top 5: 0.360, 0.312, 0.306, 0.306, 0.259
We then statistically model the distribution of problematic ads across our panel (RQ2). Our results show that problematic content makes up a relatively small fraction of all ads our users see on Facebook---a median of 10\%---but a subset of our panel is exposed to problematic ads over three times more often than the median participant. Looking at which participants tend to receive more problematic ads, we find that participants who are older are more likely to see deceptive and clickbait ads, and those who are Black are also more likely to see clickbait ads. Men are additionally more likely to see financial ads, a complex category that is (i) considered sensitive by U.S. regulation and Facebook policy, as it may include offers for exploitative financial products, (ii) disliked by participants more than neutral ads, but (iii) which also may include beneficial financial products.% and/or are men are more likely to see problematic ads.

Finally, we investigate
%Because we have access to targeting 
%data as well, we also characterize 
the extent to which the advertisers and the platform personalization algorithms are responsible for these biases (RQ3).%, depending on how narrow of a targeting criteria the advertiser originally specified. 
We find that certain categories of ads (e.g., opportunity ads and ads for sensitive topics) tend to be much more narrowly targeted than neutral ads, suggesting that advertisers carefully choose which users are eligible to see these ads.  On the other hand, we identify a subset of ads that are not targeted at all (i.e., the advertisers make all adult U.S. users eligible to see the ad), and find that demographic skews still persist for ads across different problematic categories. %For example, we find that older users receive a much higher fraction of problematic ads overall, even when the advertiser did not use any targeting options.  Similarly, we find that Hispanic users are shown a higher fraction of deceptive ads, and women see fewer financial ads. 
Together, our results shed light on users' overall ad experiences on a major platform, and illuminate disparities in those experiences caused by a combination of advertiser choices and platform algorithms.%how some users receive disparate exposure to problematic content, and the role of ad delivery optimization in the distribution of problematic content. 
%how platforms, in an effort to find relevant audiences for each ad, expose problematic content to select groups of users, and that the users themselves dislike this content.

%Taken together, our results identify potential biases in and negative outcomes of targeted advertising, and suggest directions for future work to understand the overall impact that large-scale ad platforms have on users.

\begin{comment}
Overall, our study provides a first deep look into the ad ``diets" of a diverse group of Facebook users, as well as those users' sentiments towards their own ads, with a focus on identifying disparity in exposure to potentially problematic advertising. 
%
Instead of focusing on one category of problematic ads, we take a broad and user-informed approach to identifying problematic ads from our dataset. 
%
Our results also provide a first look into how problematic ads can focus on a minority of users, who might be exposed to it repeatedly, and how personalization algorithms might be responsible for this disparity. 
%

The remainder of this paper is organized as follows. 
%
Section~\ref{sec:background} provides an overview of background material and a survey of related work.
%
Section~\ref{sec:meth} gives more details on how we recruited our panel and collected ad data, and Section~\ref{sec:categorizing} describes how we developed our codebook and classified ads.
%
Section~\ref{sec:results} presents our analysis, focusing on our three key research questions on users' ad experiences.
%
Section~\ref{sec:discussion} provides a concluding discussion.
\end{comment}

%% file: related.tex
%!TEX root = main.tex
\section{Background and Related Work}\label{sec:background}
Below, we provide background on targeted advertising, and discuss prior work on measuring skews in ad delivery, as well as users' experiences with problematic digital advertising.

% Below, we provide background on targeted advertising, and prior work measuring algorithmic skews in ad delivery. We also summarize prior social scientific work on the harms of problematic mass media broadly, as well as users' experiences with problematic digital advertising.

\subsection{Online Advertising}
\label{subsec:advertising_architecture}
Online advertising, and in particular, {\it targeted} advertising, supports much of the modern Internet's business model. 
Targeting ads to particular users can be an effective way to show content to the most relevant audiences. However, the data used in targeting is privacy-sensitive (e.g., \cite{ur2012smart,jung2017influence,mcdonald2010americans,samat2017raise}) and the targeting process can lead to discrimination (e.g.,~\cite{sweeney2013discrimination,lambrecht2019algorithmic}). 

Platforms such as Facebook and Twitter rely on inferring user's interests, and providing advertisers with an interface for these interests, to enable precise targeting of ads.
In addition to interests and behaviors, they also enable targeting by demographics (e.g., age or gender); personally-identifiable information (e.g., users' email addresses), often called ``custom'' audiences~\cite{fbcustomaudiences}; and even ``lookalike'' audiences that are able to expand a list of uploaded contacts by finding other users who have similar characteristics~\cite{facebooklookalike}.
The delivery of targeted online ads can be broken into two phases~\cite{ali2019discrimination}: {\em ad targeting} and {\em ad delivery}.
In ad targeting, the advertiser uses the targeting features described above to define an {\em eligible audience}, and specifies the ad's budget and optimization goal.
In ad delivery, the platform must decide which users in the eligible audience are actually shown the ad (the {\em actual audience}). Historically, platforms used different auction mechanisms to make this selection~\cite{edelman2007internet}, but today, platforms use sophisticated algorithms that try to subsidize ads that have high ``relevance'' to specific users~\cite{FacebookAdAuctions}.

Prior work has found that discriminatory digital advertising can result due to discriminatory targeting by advertisers~\cite{speicher2018potential,kingsley2020auditing} and discriminatory delivery by platforms, even when the advertiser might not have intended it~\cite{ali2019discrimination,ali-2019-arbiters,sapiezynski2022algorithms}.
The latter can be the result of relevance algorithms significantly skewing the the actual audience such that this audience is very different from the eligible audience an advertiser intended to reach. As a result, Facebook in particular has implemented novel systems in response to legislative pressure~\cite{facebookvrs} to minimize variance between eligible and actual audiences, in an effort to ensure fairer delivery of particular ads.
%
% Audits of Facebook's advertising platform enumerate the ways in which advertisers can knowingly and unknowingly engage in problematic ad targeting and delivery  \cite{speicher2018potential,kingsley2020auditing}. Evidence of discriminatory targeting practices have emerged \cite{sweeney2013discrimination,sapiezynski2022algorithms,lambrecht2019algorithmic}; for instance, differences in branded content on Facebook news feeds were found across genders and age groups in the Netherlands 
%(e.g., men were more likely to be exposed to ads with cars and electronics, while women were more likely to see health-related branded content)\cite{bol2020vulnerability}. Additionally, prior work has also observed how delivery optimization from platforms creates distribution bias in ways the original advertiser did not intend \cite{ali2019discrimination}. 

\subsection{Problematic Media}
Communication and psychology literature have long explored how traditional mass media (e.g., print, TV, radio) expose consumers to problematic content. Social science theories such as cultivation theory (how exposure to content may influence people's thoughts and behaviors~\cite{potter1993cultivation}) and agenda-setting theory (how content can be used to shape and filter a consumers' reality~\cite{shaw1979agenda}) posit ways in which harmful media can produce negative outcomes for consumers. Empirical observations under these frameworks include how violent media teaches violent behaviors (e.g., \cite{smith1998harmful}); 
%how drug usage in media promotes drug usage in viewers (e.g. \cite{wakefield2003role}); 
how bigoted media reinforces prejudice (e.g., \cite{vidmar1974archie,browne1999television}); and how exposure to idealized body images can lead to body image issues (e.g., \cite{van2007dieting,posavac1998exposure,aubrey2006exposure}). The ability to target mass media advertisements to specific audiences, however, is limited.

Online ads are another form of potentially problematic media. % content which internet users have little choice in consuming. 
%In this work, we draw on the foundation of prior social science works to investigate how the presence of advertisements, which are increasingly highly targeted to individual characteristics, may be problematic to users' online experiences~\cite{milano2021epistemic}. % Given this, we look to ads with characteristics that may pose harm to those exposed. Not only may the content be harmful visually in the same way traditional media can be, but, due to their digital and interactive nature, ads have the possibility of posing additional harms to users who click on them should the source be malevolent. 
Investigating the potential for advertising to expose users to problematic content is particularly important, since ad platforms often self-regulate, and set their own policies to define which advertising content they do or do not allow on their sites~\cite{fb2,google1}.
%Online websites and platforms set policies to define the content, including advertising content, that they do not allow on their sites~\cite{fb2,google1}. 
%
These policies are often updated at the platform's discretion, or in response to the changing landscape of problematic content%.
%
%For instance, social media platforms expanded their definitions of harmful content following the proliferation of misinformation related to the COVID-19 pandemic in 2020
~\cite{baker2020covid19}. To enforce these policies on all types of content, platforms use a combination of automatic and manual approaches \cite{jhaver2019human}.
But despite policies and detection tools that aim to limit problematic content, ads that users find problematic still have a significant presence on popular sites~\cite{zeng2020bad} due to both policy inadequacies~\cite{ohchr1,madio2021content} and technical challenges~\cite{sculley2011detecting}. We %synthesize prior work on harmful media and online advertising to 
investigate how the presence of ads, which are increasingly highly targeted to individual characteristics, may be problematic~\cite{milano2021epistemic}.
% The lack of uniformity, agreement, and transparency in the quality control of ads across platforms warrants a deeper look into how to identify problematic ads. 
%
%Though, in general, the way platforms moderate content is controversial \cite{ohchr1,madio2021content}.
%Researchers have described the challenge of detecting ads with adversarial content \cite{sculley2011detecting}. 

\subsection{User Experiences with Problematic Ads} 
Users of online platforms have been shown to dislike ads in general \cite{zeng2021makes,goldstein2014economic}, with some employing tools like ad blockers to browse the web without the obstruction of ads \cite{statista1}.
% Outside of the general nuisance ads present,
Recent work has investigated why users dislike online ads; Zeng et al. develop a taxonomy on what users think are the worst qualities of ads~\cite{zeng2021makes}, finding that people are particularly likely to dislike ads described as %have expanded our understanding of the nuance in users' dislike towards ads . In two separate surveys, they 1) gather open-ended responses on participants' opinions of ads, and 2) use those responses to generate opinion labels, which they ask participants to label ads with. This taxonomy of bad ad labels has revealed what users think are the worst qualities in ads; for instance, clusters of ads that were rated the lowest were described as 
``deceptive," ``clickbait,", ``ugly," and ``politicized."
% However, they found low consensus in people's descriptions of individual ads, %that few ads had high agreement on opinion labels, 
%suggesting that people think differently about what makes ads problematic.
%bad, and may disagree about whether a particular ad is problematic.
People struggle to identify deceptive ads~\cite{xie2011susceptible}, which can lead to harmful outcomes like software attacks~\cite{nelms2016towards,xie2015disentangling}. Those who suffer from certain mental health disorders or trauma may also experience negative psychological and physical consequences from ads that target these conditions~\cite{gak2022distressing}.
%\medskip
%Our work lies at the intersection of prior qualitative and quantitative work. 

\parb{Our contributions.} We build on prior qualitative work by Zeng et al.~\cite{zeng2021makes}, and use their taxonomy to assess people's sentiments toward their own ads.
We further use these sentiments, combined with rigorous coding, to identify novel categories of ads perceived, more specifically, as {\it problematic}.
We also extend prior quantitative work, such as Ali et al.~\cite{ali2019discrimination}, and measure ad delivery's role in creating disparities in exposure to problematic advertising.
We further show ad delivery biases are not limited to ads created by researchers~\cite{ali2019discrimination}, and extend to real problematic ads on the platform.
To our knowledge, ours is the first study to look at targeting and personalization of problematic ads to actual users.

%% file: methods.tex
\section{Methodology}
\label{sec:meth}
% In order to investigate real-world user experiences and inequities in problematic advertisements, we longitudinally collect ads from a panel of 132 Facebook users and measure their sentiment toward the ads they receive.
Below, we describe our methods for recruiting a diverse and demographically balanced panel and for collecting the desktop ads our participants are shown by Facebook.
%
%In the following section, we describe how we develop a taxonomy of participants' advertisements, and survey our participants' sentiments toward the ads they receive.% about their ads 
%
%
%On a high level, our study design includes (a) recruiting a diverse and balanced panel of Facebook users, (b) observing all of the ads they are shown by Facebook through our browser plugin, (c) manually annotating those ads into a fixed set of categories, and finally (d) surveying our panel monthly to understand their perceptions of our categories.
%
%Our motivation behind this design is to broadly observe experiences of real users, and more specifically, to understand the difference in exposure to advertisements that are potentially harmful to their consumers.
%
%In this section, we describe how we completed each of these four steps in detail.

\subsection{Panel Recruitment}
% removed: inequities in real-world...
%To examine real-world experiences with problematic ads, we needed to recruit a diverse panel of Facebook users from whom we could collect and assess their advertisements. %Our goal in recruiting a panel of users was to observe the advertising experience for a diverse population of Facebook users.
%
We recruited our panel of Facebook users from two sources: by listing tasks on Prolific, an online crowd-work and survey platform, and by advertising on Facebook. 

Participants were screened via a short survey.\footnote{Prolific participants were compensated with a base pay of \$8.04 per hour for completing the screening survey while those recruited via Facebook advertisements were not compensated for the screening survey as there is no mechanism to do so. Demographics which we initially struggled to recruit were offered marginally more compensation. The survey took a median of 6 minutes and 9 seconds to complete.} Our criteria to be eligible for the study were that participants must (1) have an active Facebook account that (2) they use for at least 10 minutes per day (3) on a desktop or laptop computer (4) via either the Google Chrome or Mozilla Firefox browsers (5) without using ad blockers or tools for anonymous browsing (e.g., Tor). Additionally, we went to significant lengths to recruit a diverse panel across select demographic variables: race and ethnicity (white; Black; Hispanic; Asian), gender (men; women), age (younger than Generation X \cite{pew2}; Generation X or older) and educational attainment (below a bachelor's degree; bachelor's degree and above).
We sought to balance our panel among all combinations of our chosen demographic variables (e.g., representation for Generation X Hispanic women with high educational attainment) but we struggled with recruitment and retention of some demographics, 
%(older people, less educated people, and --- to a lesser extent --- Hispanic people), 
partly due to the distribution of users who participate in online studies or use the platforms we recruited on \cite{prolific1,tang2022well,pew1}.
We made a continuous effort to balance our sample by accepting participants on a rolling basis and not screening in those with demographics we were saturated with.
Table~\ref{tab:panel_demographics} shows the ultimate demographic breakdown of our participants. 

Unfortunately, while all participants were screened based on their Facebook usage, not all users contributed a significant number of ads during the 3 month study period. Of the 184 participants originally enrolled in the study, 132 were {\em active} participants, which we define as those who contributed at least 30 ads (on average 10 per month) over the course of the three months of their participation in the study.

\input{tables/demographics_table}

\subsection{Data Collection} \label{sec:data_collection}
\smallskip
\noindent\textbf{Logged Data.} Our study collected the ads that were shown to our participants on their Facebook news feeds while using Facebook on a desktop computer over a 3 month period. In order to collect our participants' ads, we used a browser extension, based on the NYU Ad Observer project~\cite{PropublicaFacebookPoliticalAds, socialmediacollector}.
We modified Ad Observer to include unique participant IDs along with the ads reported to our server, and we introduced an additional ``Surveys'' tab that serves participants monthly surveys to collect their sentiments for their individual ads.
Across all of our recruited participants, we collected 165,556 impressions to 88,509 unique ads. Repeat impressions of ads are relatively sparse in our data---a median of twice per ad per participant---and only 5.33\% of our ads are shown more than 3 times to a participant.

\smallskip
\noindent\textbf{Targeting Data.} We also collected ad targeting information provided by Facebook through its ``Why am I seeing this?" API~\cite{FacebookWAIST}, which reveals information about how the advertiser selected their target audience~\cite{andreou-2018-explanations}. While prior work has shown that Facebook's targeting explanations can be incomplete, and include only one targeting criteria in each ad explanation~\cite{andreou-2018-explanations}, we find empirically that the system has changed since. We also observe differences between the summarized targeting data which is shown on the user interface, and what is reported through the  API.
Our data includes several instances of multiple targeting criteria---62.7\% of ads in our data with interest targeting include more than one interest.

%
% select count(*) from observations inner join users on users.instid=observations.instid and users.pid in (<tuple of active pids>);

\smallskip
\noindent\textbf{Survey Data.} Every month, we prepared a survey that assesses participant sentiments toward the ads they saw on Facebook during the prior month. Specifically, for each ad that we showed to a user in the survey, we asked them:
%Our first question asked, {\bf Q1.} ``How would you describe the advertised product/offer's relevance to you?'' with a 5-point Likert scale of ``Completely Irrelevant" to ``Completely Relevant." We do not use this question in our analysis.
%We then ask, {\bf Q2.} 
``Which of the following, if any, describe your reasons for {\it disliking} this ad?" and present the following non-mutually exclusive answer choices:
\begin{packed_itemize}
    \item It is irrelevant to me, or does not contain interesting information.
    \item I do not like the design of the ad.
    \item It contains clickbait, sensationalized, or shocking content.
    \item I do not trust this ad, it seems like a scam.
    \item I dislike the advertiser.
    \item I dislike the type of product being advertised.
    \item I find the content uncomfortable, offensive, or repulsive.
    \item I dislike the political nature of the ad.
    \item I find the ad pushy or it causes me to feel anxious.
    \item I cannot tell what is being advertised.
    \item I do not dislike this ad.
\end{packed_itemize}

\noindent We then ask: ``Which of the following, if any, describe your reasons for {\it liking} this ad?" and present the following non-mutually exclusive answer choices:
\begin{packed_itemize}
    \item The content is engaging, clever or amusing.
    \item It is well designed or eye-catching.
    \item I am interested in what is being advertised.
    \item It is clear what product the ad is selling.
    \item I trust the ad, it looks authentic or trustworthy.
    \item I trust the advertiser.
    \item It is useful, interesting, or informative.
    \item It clearly looks like an ad and can be filtered out.
    \item I do not like this ad.
\end{packed_itemize}

% Answer choices for these questions are drawn from Zeng et al.'s taxonomy of reasons users like or dislike ads~\cite{zeng2021makes}, with the exception of one item.
Answer choices for these questions are drawn from Zeng et al.'s taxonomy of reasons for users' like or dislike of ads~\cite{zeng2021makes}, with the exception of one item.
In a small pilot version of this survey, in which we allowed participants to also provide free-text answers of their reasons for liking and disliking Facebook ads with 300 respondents, we identified an additional reason for liking an ad, “This ad is filterable”, so we included it to capture a broader spectrum of reasons users like ads. 

We survey participants about at most %35 ads per month --- 
5 ads from each of our seven ad categories (Section~\ref{sec:categorizing}). We limit the monthly surveys to up to 35 ads each so that it did not become prohibitively long (more than 20 minutes) for participants to complete. %\elissa{talk about sensitive being broken out b/c distribution here}: neutral, healthcare, opportunity, clickbait, deceptive, potentially prohibited, sensitive, financial. 
%Participants access surveys through our extension's surveys tab.
%
%
%We send participants an email notification when a new survey is available within the plugin's ``Surveys" tab.
%

\smallskip
\noindent\textbf{Study Deployment.} We began data collection in November 2021, with participants recruited on a rolling basis. Each participant was a part of our study for three months.
The final participant completed the study in September 2022.
We compensated our participants by paying them up to \$60: \$5 when they signed up, \$15 for each month they kept the plugin installed and completed the monthly sentiment survey, and upon completing all three months of the study, they were rewarded with a \$10 bonus payment.
Those participants who dropped out of our study were compensated using the scheme above based on how long they did participate. Since we deployed surveys directly through our extension, we were not able to assess average time of completion, but pilot tests of the survey averaged a completion time of about 15 minutes.

\subsection{Analysis}
\label{sec:meth:analysis}

\noindent Here, we describe the quantitative methods we employ to analyze survey responses, logged ad observations, and ad targeting data. We limit all our analyses to the  32,587 ads that we annotated (see Section~\ref{sec:categorizing}), and to our list of active participants (Table~\ref{tab:panel_demographics}).

\smallskip
\noindent\textbf{RQ1.} For survey responses, we use  Chi-squared ($\chi^2$) tests for equality of proportions to compare rates of ad dislike.
We also report Cohen's $\omega$ as the effect size of the Chi-squared tests to characterize the scale of differences.
As a general guideline, $\omega=0.1$ is considered a small effect, 0.3 is a medium effect, and 0.5 and above is considered a large effect~\cite{cohen2013statistical}.
We examine the association between the reasons for dislike mentioned in the surveys and the ad type through mixed-effects logistic regression models.
%For each ad type, we model the ad type (a boolean) as a dependent variable, and each reason for dislike from Q3 (also boolean) as an independent variable.
To control for variance in participants' individual preferences, we include a random effect term for each participant.
In line with statistical best practice~\cite{gelman2012we}, we do not correct our regression models as each model represents a purely complementary comparison (e.g., contains a distinct dependent variable).

\noindent\textbf{RQ2.} To understand disparities in the distribution of ad types, we treat number of ad types observed for each participant as a frequency distribution.
To quantify inequality in this distribution, we compute skewness~\cite{wiki_skewness}, a measure of asymmetry for a probability distribution, computed via its third standardized moment.
% Concretely, skewness of a random variable $X$ with mean $\mu$ and standard deviation $\sigma$, given $N$ observations is: $$\frac{\sum_{i=1}^N (X_i - \mu)^3/N}{\sigma^3}$$
A positive skew implies a distribution with a long right tail, while a negative skew means the left tail is longer.
We also compute the Gini coefficient~\cite{wiki_gini} to measure inequalities across participants.
% If each participants $i$ sees a certain ad type $x_i$ times, and all participants see it an average $\bar{x}$ times, the Gini coefficient is computed by comparing differences over all pairs of participants:
% $$\text{Gini} = \frac{\sum_{i=1}^n{\sum_{j=1}^n} |x_i - x_j|}{2 n^2 \bar{x}}$$
%
To understand inequities between demographic groups, we use linear regression models to model the fraction of each ad category in participants' ad diet, as a function of their demographics.

\noindent\textbf{RQ3.} To disentangle ad delivery's influence from ad targeting in our observations, we use the advertising interface to obtain audience size estimates for each ad. Concretely, we query Facebook's advertising API for monthly ``reach'' estimates for the targeting specifications of every ad in our dataset. Note that these estimates are not accessible for ads that use Custom Audiences (CAs), such as phone number uploads or cookie re-targeting; those are only known to the owners of these CAs.
We use linear regressions similar to RQ2 to identify differences between demographic groups that appear due to the platform's ad delivery practices.
%\elissa{describe the overall analysis methods here per RQ/section...I'll come back later}

\subsection{Ethics}
Given the sensitivity of the data we were collecting, we took care to follow best practices, maximizing beneficence while minimizing harm to our participating users and Facebook itself.
First, our research project was approved by our institution's Institutional Review Board (IRB). %\footnote{IRB approval number not included for anonymity.} 
Second, we collected the minimal data on our participating users necessary to conduct the study; we only collected personally-identifiable information where necessary to facilitate payments, and we used unique, random identifiers for all survey responses and ads collected.
Third, we controlled access to the uploaded pseudonymous data to just the research team, and we do not plan on making this data generally available to protect the privacy of our users.
Finally, we minimized the harm to advertisers and Facebook itself by not causing any ad impressions that would not have otherwise occurred; the only additional requests to Facebook were to fetch the targeting specifications, and to later retrieve audience sizes of these specifications.
%provided in the ``Why am I seeing this ad?'' pop-up.

% ref: https://www.facebook.com/terms/, Section 3.2.3
While Facebook prohibits collection of data using automated means in its terms of service (ToS), we argue that the public benefits of our work outweigh the risks posed to Facebook.
Further, violating ToS by scraping content that is otherwise available through non-automated means is not considered a violation of the U.S. Computer Fraud and Abuse Act~\cite{2021van}. 
Platforms, however, reserve the right to ban users who scrape or have done so in the past.

\section{Categorizing ads}\label{sec:categorizing}
% Average # of annotated ads: select avg(cnt) from (select pid, count(*) as cnt from coding left join pid_adid on coding.id=pid_adid.id where pid in <active_pids_tuple> group by pid) as temp;
In order to evaluate whether there are inequities in participants' exposure to problematic ads, we first evaluate which of our collected ads are problematic. 
To do so, we develop a codebook to categorize the ads our participants see, and then use that codebook to annotate a significant subset of their ads.
%
%First, we use rigorous qualitative methods and prior literature to develop a codebook of different categories of potentially problematic ads. 
%
%Second, using this codebook, four coders annotated a random sample of 32,587 (39.0\%) of the advertisements our participants were shown. %On average, we annotate 546 ads for each active participant. 
%
%Third, we use carefully-constructed quantitative surveys to assess participant sentiment toward a random sample of ads across the seven categories. 
%
%We do this to identify the categories of ads the participants find problematic, which we later analyze for biases (Section~\ref{sec:meth:analysis}). %Here, we describe each of our advertisement assessment steps in detail.

\subsection{Creating the codebook}
% To understand the content that our participants see, we create a codebook to categorize each participant's ads into a fixed set of categories. 
%
We use a combination of inductive qualitative coding~\cite{thomas2003general,chandra2019inductive},
%(looking at ads to identify common themes)
and deductive analysis~\cite{azungah2018qualitative} of prior work and platform policies
%(literature review of prior work as well as government and platform policies)
to develop a robust categorization of participant ads.

To create our initial inductive categorization of Facebook ads, we conducted pilot data collection with 7 participants, collecting their ads with our browser extension between June and July 2021. 
We then cross-referenced our initial codebook with platform and governmental policies and empirical research to develop our final ad categories. 
Our categorization particularly focuses on capturing problematic ads, though we also make sure our codebook captures content that users might find unproblematic, such as products, events, or local businesses. 
%We further validated our categories on the 3700 ads annotated during the first month of our full data collection, refining how we coded each category as needed. 
Below, we define our categories, describe how we reason about them, and provide examples from our dataset.

% Each subheading here should answer 1: why is this bad? if not bad then important, 2: cite other people who have thought about this to establish that we don't pull it out of our ass
\parc{\deceptive{}:} Fraudulent offers, potential scams, false or misleading claims, predatory business practices. {\it Examples:} Guaranteed monthly income,
% fat burning pills,
sign-up flows for personal information (“clickfunnels”), non-descript offers with requests for direct messages.

Deceptive advertising and its breadth is notoriously hard to capture (see, e.g., a review of definitions~\cite{ford2019data} and a diversity of FTC reports on the subject~\cite{ftc_challenging}).
Therefore we define this code broadly, to be able to capture multiple forms of deceptive and scam content.
We categorize financial and personal information scams, fraudulent offers,  and a diverse array of misleading content as \deceptive{}. Many aspects that we cover in this definition are covered by Facebook's policies for unacceptable business practices~\cite{fb_unacceptable_business}, unrealistic outcomes~\cite{fb_unrealistic_outcomes}, and broadly under the platform's deceptive content policy~\cite{fbadstandards}.
Prior work has documented deceptive ads in contexts such as malicious web advertising \cite{li2012knowing}, social engineering attacks \cite{nelms2016towards}, and distributing malware \cite{rastogi2016these,zarras2014dark}.

%\smallskip
%\noindent {\it Examples:} Guaranteed monthly income, fat burning pills, sign-up flows for personal information (“clickfunnels”), non-descript offers with requests for direct messages.

\parc{\clickbait{}:} Ads that omit information to entice users, are unclear about the advertised product, or contain sensational, loud, or dense content. {\it Examples:} Provocative news headlines, celebrity gossip, incomplete offers (``Click to find out'').

Prior work has documented how clickbait ads are attention grabbing by being unclear, and do not live up to users' expectations~\cite{zeng2021makes,pengnate2021effects}.
It has also been found to waste users' time~\cite{scott2021you}, contain provocative content \cite{pengnate2016measuring}, and act as a vehicle for misinformation \cite{pengnate2016measuring,gdi1,zeng2021polls}.
Facebook's policies also recognize the misleading and annoying nature of clickbait, and they enforce policies to reduce exposure to such content~\cite{fb1,fbaddemote}.

%\smallskip
%\noindent {\it Examples:} Provocative news headlines, celebrity gossip, incomplete offers (``Click to find out").

\parc{\prohibited{}:} Ads that may not be allowed on the platform according to Facebook's prohibited content policies. {\it Examples:} Tobacco, drugs, unsafe dietary supplements, multi-level marketing, weapons.

% Ads that are classified as potentially prohibited are those that appear to violate ad platforms' own policies on what kind of ads are allowed on the platform.

Facebook's policies prohibit several types of ads~\cite{fbadstandards}, including but not limited to ads for tobacco, adult content, body parts, payday loans, and multi-level marketing.
Ads that pose a security threat to users, such as spyware or malware, non-functional landing pages, and efforts circumventing review systems, are also prohibited~\cite{fb2}.
Even with an extensive policy, Facebook's ability to accurately detect content and enforce policies is limited (see, e.g., prior work documenting challenges in detection and enforcement of  political advertising policies~\cite{edelson2020security,le2022audit}). We therefore code for ads whose content match any of Facebook's prohibitive policies. We note that only Facebook can enforce these policies -- therefore we refer to our annotations as {\it potentially} prohibited.

%\smallskip
%\noindent {\it Examples:} Tobacco, drugs, unsafe dietary supplements, multi-level marketing, weapons.
%
% For instance, recent analyses of Facebook find that the advertising platform contains weaknesses that allow political ads to go undisclosed by advertisers \cite{edelson2020security}, and that Facebook was largely inaccurate at detecting political ads on the platform themselves \cite{le2022audit}. 
%
%Our analysis of the ad landscape of Facebook users will shed light on whether certain users are being exposed to harmful content that, according to policy, they shouldn't be. We compare ads against Facebook's guidelines for prohibited ad content to build this category.

\parc{\sensitive{}:} Ads that fall under Facebook's content-specific restrictions policy~\cite{fbadstandards}: such content isn't prohibited but, given its sensitive nature, it must comply with additional guidelines, including written permissions and certifications. {\it Examples -- Sensitive: Financial:} Credit cards, loans, mortgage financing.
{\it Examples -- Sensitive: Other:} Weight loss programs, online mental health prescription services, online slot machines.

Facebook subjects ads for sensitive topics 
%such as weight loss, gambling, alcohol, over-the-counter and prescription drugs, online pharmacies, and financial products 
to additional scrutiny on their content and targeting practices~\cite{fbadstandards}.
For example, ads for weight loss programs can only be targeted to people at least 18 years or older, financial advertisers must provide authorization by regulatory authorities, and online pharmacies require an additional certification~\cite{fb_pharmacies}.
Within Sensitive ads, we find an increased prevalence (more than two-thirds) for Financial ads, so we break this code into two sub-codes --- \sensitivefinancial{} and \sensitiveother{}.

In addition to platform policies, sensitive ads closely relate to prior work on content that targets user's vulnerabilities~\cite{gak2022distressing,algorithmsoftrauma} --- such content may be benign to some users but may foster negative thoughts or behaviors for others~\cite{milano2021epistemic,sheknows1}.
Gak et al.~\cite{gak2022distressing}, for instance, found that among people with a history of unhealthy body stigmatization, dieting, or eating disorders, being targeted with weight-loss-related ads had negative emotional and physical outcomes.

\parc{\opportunity{}:} Ads that present any employment, housing, or educational opportunity to users. {\it Examples:} Degree programs, jobs or gig-work, fellowships, scholarships.

We coded for ads that displayed opportunities for users, such as a job or gig, higher education, or apartments and homes for sale. Facebook's own policies prohibit discrimination in targeting of opportunities, or advertising fraudulent or misleading opportunities~\cite{fb_jobs_policy}. Further, cases of discrimination in the delivery of online opportunity ads \cite{datta2018discrimination,kingsley2020auditing,ali2019discrimination} led us to code these ads to examine their distribution among our participants. 

%\smallskip \noindent
% {\it Examples:} Degree programs, jobs or gig-work, fellowships, scholarships.

\parc{\healthcare{}:} Ads that contain products, services or messages related to healthcare, fitness, mental and physical wellness. {\it Examples:} Medical devices, gym equipment, public health announcements, fitness programs, health insurance.

%Healthcare ads include a wide array of healthcare-related content, from fitness programs to health insurance services.
%
We find a wide array of healthcare-related ads that are broader than the content covered by Facebook's content-specific restrictions (Sensitive), and we use a separate code to capture such content. These ads are diverse in nature, ranging from helpful to possibly problematic.
% are particularly regulated on online \cite{fb1,google1} and offline \cite{fda1} platforms. Should healthcare ads be found to contain problematic features, they pose direct risks for physical harm to users (e.g. ingestion of unsafe dietary supplements, following unauthorized medical advice).

%\smallskip \noindent
 %{\it Examples:} Medical devices, gym equipment, public health announcements, fitness programs, health insurance.

\parc{\political{}:} Ads that contain any overt references to political subject matters. {\it Examples:} Political campaign ads, petitions for political causes.

While we initially coded for political ads, we exclude them from our analysis. We consider ads for political content to be outside of our scope for this study due to challenges in measuring user perceptions of political ads~\cite{sosnovik2021understanding}; further, problematic content \cite{zeng2021polls}, delivery~\cite{ali-2019-arbiters} and policy \cite{le2022audit} surrounding political ads are well-addressed in recent prior work.

%Platforms struggle to detect political ads and hold them to their policy standards \cite{le2022audit}, enabling the use of problematic tactics like deception  \cite{zeng2021polls}. While not generalizing all political ads as problematic, we keep them in a separate category.

%\smallskip \noindent
 %{\it Examples:} Political campaign ads, petitions for political causes.

\parc{\neutral{}:} Every-day products, services or apolitical news. {\it Examples:} Sales, product deals, local events. Further, ads not classified as any of the other categories are considered neutral.

%\smallskip \noindent
% {\it Examples:} Sales, product deals, local events.

%Neutral ads are ads which we, the researchers associated with this work, agreed gave us no reason to believe they would be problematic for our participants through deception, clickbait, displaying prohibited content, or touching on any well-researched mental health disorders that may be sensitive topics. Within this category are ads for clothing products, household services, or local community events. These ads also did not present any clear opportunities to users, nor did they fit into our categories of Healthcare, Political, or Financial. \angelica{ali, did we ever double code benign with anything, like benign healthcare? i'm remembering no, but if we did i'll take this sentence out}

%
\medskip

%Since our codebook is inductive and built from a subset of our data, we also encounter and categorize ad types we might not have initially expected, such as ads for lawsuits and political content (our data collection ran outside of any major US political events).
%
\noindent The prevalence of each category in our annotated data is shown in Table~\ref{tab:codebook}. Figure~\ref{fig:ad_images} also shows concrete examples of each category. We leave a small fraction of ads (122, 0.41\%) in our dataset uncategorized because they do not fit into our codebook, but are also not benign; often, these are potentially deceptive offers which we are unable to verify.
Since some of our participants are recruited from Facebook, we observe an increased prevalence of research-study-related % and market research 
ads (2558, 7.85\%). We use an auxiliary code ``Study'' to annotate all such ads, and remove them from all subsequent analyses.

In our annotation, we allowed for double-coding when an ad fell into two or more categories (e.g., an unclear ad for ``5 Steps My Clients Use to Overcome Anxiety'' falls into both \healthcare{} and \clickbait{}). However, we do not allow multiple codes when an ad is categorized as \neutral{}. 

\begin{table}[h]
\centering
\input{tables/codebook_table_condensed.tex}
\caption{Prevalence of each code in our annotated dataset.}
\label{tab:codebook}
\end{table}

\subsection{Coding ads} 
Across all of our recruited participants, we collected 165,556 impressions to 88,509 unique ads. 
Out of these, 83,507 (94.3\%) ads and 156,213 (94.3\%) impressions were contributed by the participants ultimately deemed active (and considered in the remainder of the study). Due to the high volume of ads, 
%we were only able to code a subset of our participants' ads. 
%
%To prepare our monthly surveys, 
we annotated a random subset of up to 200 ads per participant per month. 
Since we repeated this sampling strategy every month for each participant, we avoid introducing time- or participant-related sampling biases to the subset of our data we annotated. 
Through this sampling process, we were able to annotate 32,587 out of our collected 88,509 ads, or $\approx$36.8\% of them. 

The authors annotated the first two months of data.
%Annotation for the first two months of the study was done by the first two authors of this study. 
%
For the remaining months, we hired two students from our institute as external annotators. 
We choose to hire annotators locally instead of crowd-workers to be able to train them to use our codebook properly and communicate in case of errors. 
The annotators were shown the ad's text and a screenshot of the ad (e.g. Figure~\ref{fig:ad_images}) during annotation tasks.

Since our annotation task consists of multiple labels and we consider agreement for more than two annotators, we use Krippendorf's Alpha with the Jaccard set distance function to evaluate agreement between annotators.
External annotators were first trained to use the codebook on a pilot task using the authors' gold standard annotations. 
Subsequently, every month, we picked a 5\% subset of the month's ads to overlap across both annotators and the first author. 
If agreement on this common subset was low ($\alpha$ $<$ 0.70), we went over discrepancies and re-calibrated our use of the codebook. 
We repeated this exercise each month to ensure annotation quality remained high. 
The final agreement on our annotated data, $\alpha=0.726$, is considered `substantial'~\cite{landis1977measurement}.

We specifically avoided using machine learning to avoid mis-labeling points in our data. 
\deceptive{} content, in particular, requires a level of investigation that would not be possible with automation.
To investigate whether an ad is indeed deceptive, annotators are asked to visit the advertised web page, look at the advertiser's Facebook page, and inspect reviews on Facebook and Better Business Bureau.
%
% Similarly, identifying \prohibited{} content requires knowledge of Facebook's content policies, which every annotator was asked to read.
%
% Moreover, our dataset was likely not sufficiently large for a machine learning-based approach to outperform hired annotators.   

\parb{Post-processing.} Finally, while we annotate multiple codes per ad for a richly described dataset, we post-process our coding to translate into one code per ad. We do this for easier interpretation of the following results (Section~\ref{sec:results}), particularly in regression analyses. In line with the severity of restrictions in Facebook's policies~\cite{fbadstandards}, we translate sets of codes to a single code in the following precedence order:\newline \prohibited{} $>$ \deceptive{} $>$ \clickbait{} $>$ \sensitive{} $>$ \opportunity{} $>$ \healthcare{} $>$ \neutral{}.

%% file: tables/demographics_table.tex
\begin{table}
\centering
\resizebox{\columnwidth}{!}{
\begin{tabular}{r  l  r  r  r  r  r} 
\multirow{2}{*}{\bf Variable} & \multirow{2}{*}{\bf Value} & \multicolumn{2}{c}{\bf Recruited} & \multicolumn{2}{c}{\bf Active} & {\bf Census}\\
 & & \multicolumn{1}{c}{\textbf{n}} & \multicolumn{1}{c}{\textbf{\%}} & \multicolumn{1}{c}{\textbf{n}} & \multicolumn{1}{c}{\textbf{\%}} & \multicolumn{1}{c}{\textbf{\%}}\vspace{.3em}\\
 \toprule
\multirow{3}{*}{\textbf{Gender}} & Female & 96 & 52.17 & 71 & 53.79 & 50.5\\
& Male & 86 & 46.74 & 59 & 44.70 & 49.5 \\
& Non-binary & 2 & 1.09 & 2 & 1.52 & --\vspace{1.2em}\\

\multirow{2}{*}{\textbf{Age}} & Younger than Gen-X & 134 & 72.83 & 88 & 66.67 & 33.6\\
& Gen-X and older & 50 & 27.17 & 44 & 33.33 & 47.8\vspace{1.2em}\\

\multirow{5}{*}{\makecell{\textbf{Race} / \\ \textbf{Ethnicity}}} & White & 105 & 57.07 & 82 & 62.12 & 75.8\\
& Latino/Hispanic & 21 & 11.41 & 16 & 12.12 & 18.9\\
& Black & 53 & 28.80 & 32 & 24.24 & 13.6\\
& Asian & 21 & 11.41 & 16 & 12.12 & 6.1\\
& Other & 3 & 1.63\ & 3 & 2.27\ & --\vspace{1.2em}\\

\multirow{2}{*}{\textbf{Education}} & Below Bachelor's & 72 & 39.13 & 51 & 38.64 & 58.5\\
& Bachelor's or above & 112 & 60.87 & 81 & 61.36 & 32.9 \\
\midrule
\textbf{Total} & & \textbf{184} & & \textbf{132} & \\
\end{tabular}
}
\caption{Demographics of panel participants.% Active participants are those who contributed at least 30 ads over the course of the study. Population percentages from the U.S. 2020 Census shown for comparison of representativeness.
}
\label{tab:panel_demographics}
\vspace{-1em}
\end{table}

%% file: tables/codebook_table_condensed.tex
%\rotatebox{90}{
{\small
% \resizebox{.8\textwidth}{!}{
\begin{tabular}{lrr}

\textbf{Code} & {\bf Count} & {\bf \%}\\
\toprule
Neutral& 20,596 & 68.52 \\
Healthcare & 3564 & 11.86 \\
Opportunity & 2267 & 7.54 \\
Sensitive: Financial & 1429 & 4.75\\
Sensitive: Other & 631 & 2.10\\
Clickbait & 1182 & 3.93\\
Deceptive & 542 & 1.80\\
Potentially Prohibited & 253 & 0.84\\
% \midrule
% {\bf Class\newline Action\newline Lawsuit} & Ads that contain information related to class action lawsuits that may be applicable to users. & 437 & 1.45 \\
Political & 263 & 0.87\\
% \bottomrule
% \vspace{1em}
\end{tabular}
}
%}

%% file: results.tex
\section{Results}\label{sec:results}
We now summarize our study's results. Section~\ref{subsec:participant_perceptions} identifies which categories of ads participants find problematic (RQ1). Section~\ref{sec:distribution_analysis} investigates the distribution of problematic ads (RQ2). Section~\ref{sec:targeting_analysis} examines the reasons for the discovered discrepancies (RQ3). 
% Section~\ref{sec:case_studies} discusses individual case studies of participants who consume the highest amount of problematic ads in our panel. 

% \parb{Preprocessing.} 

\subsection{What do participants find problematic?}
\label{subsec:participant_perceptions}

\begin{figure}[th!]
  \centering
  \includegraphics[width=\linewidth]{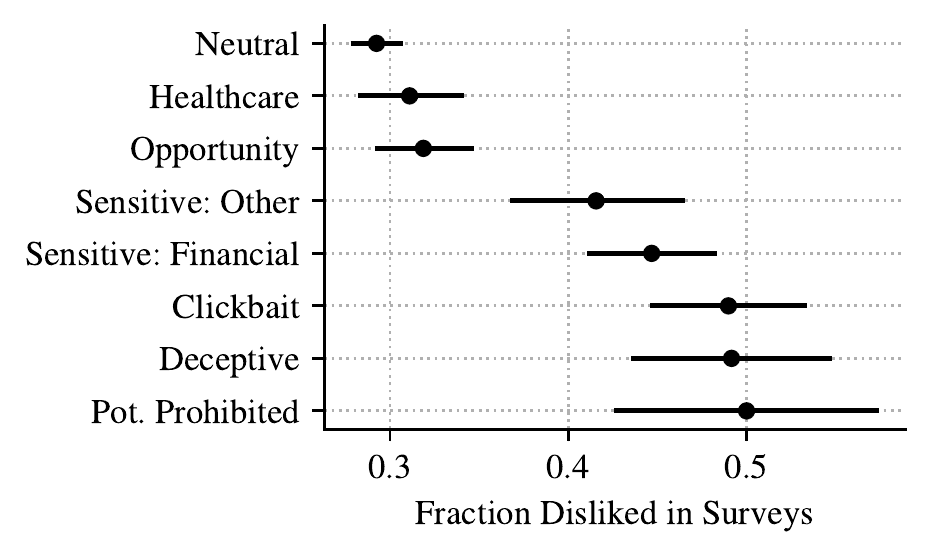}
\caption{Fraction of responses where participants showed dislike for an ad category (i.e., chose ``I do not like this ad" in the survey). 95\% confidence intervals for (binomial) proportions are estimated via normal approximation.}
\label{fig:participant_perceptions}
\label{fig:survey_responses}
\end{figure}

To evaluate whether our participants found certain ad categories problematic, we first examine general dislike: whether participants dislike a higher fraction of particular ads. We then evaluate reasons for disliking: whether participants have different reasons for disliking each category in our codebook.
Specifically, to evaluate general dislike, we use $\chi^2$ proportion tests to evaluate differences in the proportion of ads in each category that participants marked as ``I do not like this ad'' in the second question of our survey (Section~\ref{sec:data_collection}).
% in Q2 on our survey (Section~\ref{sec:data_collection}).

Figure \ref{fig:survey_responses} shows the fraction of responses, for each category, that were disliked by participants.
%
% \elissa{not exactly clear on this phrasing, I think you mean these \% are the \% of ads surveyed that were disliked, tried to rephrase accordingly.}
Across our surveys, participants reported disliking nearly half of the ads we had classified as \clickbait{}  (48.98\%), \deceptive{} (49.16\%), and \prohibited{} (50\%). Participants reported disliking 43.58\% of the ads we coded as \sensitive, % they We observe here that Clickbait, Deceptive and Potentially Prohibited ads are disliked nearly 50\% of the times they were presented in surveys, while Sensitive ad types are disliked in over 40\% of responses. 
while \neutral{}, \healthcare{} and \opportunity{} ads were disliked less: 29.24\%, 31.09\%, and 31.87\%, respectively.
%
\input{tables/dislike_reasons_regression.tex}

%An omnibus $\chi^2$ proportion test finds 
These differences across ad categories are significant ($p<0.001$, omnibus $\chi^2=186.25$; $\omega=0.15$).
In a series of pair-wise $\chi^2$ proportion tests comparing each of our coded categories with Neutral, with Benjamini \& Hochberg correction~\cite{benjamini1995controlling},
we observe that \textsf{Potentially Prohibited}, \textsf{Deceptive}, \textsf{Clickbait}, and both types of \textsf{Sensitive} ads (\textsf{Financial} and \textsf{Other})
are all disliked significantly more than \textsf{Neutral} ads ($p < 0.001$, $\chi^2 > 24$; $0.07 \leq \omega \leq 0.13$).
\textsf{Opportunity} ($p=0.121$, $\chi^2=2.60$; $\omega<0.10$) and \textsf{Healthcare} ($p=0.28$, $\chi^2=1.14$; $\omega<0.10$) ads, on the other hand, are not significantly more or less disliked than \textsf{Neutral} ads.
To identify whether any of the ad categories are disliked more than each other (rather than just more than \neutral{}) we conduct an additional series of pair-wise corrected tests, comparing differences between sequential ad categories (e.g., comparing \prohibited{}, the most disliked category, with \deceptive{}, the next most disliked).
This testing finds only one significant difference, between \sensitiveother{} and \opportunity{} %Further pair-wise proportion testing between ad types finds differences between Potentially Prohibited, Deceptive, Clickbait and Sensitive ads to be insignificant ($p > 0.1$); differences between Opportunity, Healthcare and Neutral are also insignificant ($p > 0.1$). We find that only the difference in dislike between Sensitive: Other and Opportunity is significant 
($p=0.003$, $\chi^2=11.34$; $\omega<0.10$). In combination, our statistical results suggest that \clickbait{}, \deceptive{}, \prohibited{}, and \sensitive{} ads form an equivalence class of potentially problematic ads.
%there are two equivalence classes in terms of participants' dislike in our codebook: Sensitive, Clickbait, Deceptive and Potentially Prohibited are ad types that are {\it highly disliked}; Healthcare and Opportunity ads, while highly personalized, are disliked at a similar rate to other neutral forms of advertising --- and are therefore {\it less disliked}.
% ========== SHOULD BE COMMENTED OUT IF PEOPLE DON'T LIKE IT
% Additionally, observing rates of like in survey responses, we find these categories are also significantly less liked than \neutral{} ads (see Appendix~\ref{sec:rates_like}).

To understand {\it why} participants dislike these ad categories, 
%---whether they have reasons for finding them problematic, beyond a general dislike---
we investigate the specific reasons they reported for disliking in the first survey question.
Table~\ref{tab:dislike_reasons} shows the odds ratios (exponentiated regression coefficients) of eight mixed-effects logistic regression models, with a random intercept for the participant. The odds ratios (O.R.) give the relative odds that an ad category was described with a certain dislike reason in survey responses, compared to the same dislike reason for our baseline (\neutral{}).
For each ad category (column), an O.R. of 1 means a given dislike reason (row) is not used to describe the ad category more often than Neutral. Values greater than 1 correspond to increased odds of participants describing that ad category with the given reason, while values less than 1 indicate lower odds.

We first observe in Table~\ref{tab:dislike_reasons} that participants are significantly more likely to describe the combined most highly disliked ad categories (``Problematic'' column) as irrelevant (O.R. = 1.34, $p=0.011$), clickbait (O.R. = 1.47, $p=0.018$) and scam (O.R. = 1.64, $p < 0.001$). Looking at the disliked categories individually, we find that \deceptive{}, \clickbait{} and \sensitive{} ads are also significantly more likely to be described as scams (all O.R. $\geq$ 1.45, $p < 0.05$).
The odds of \sensitiveother{} ads, in particular, being described as scams are more than twice the odds of \neutral{} ads being described as scams (O.R. = 2.08, $p=0.001$).
%---suggesting that participants do not trust ads for e.g., gambling and dieting, which fall in the Sensitive: Other category. 
Also for these ads, participants' odds of disliking the advertiser (O.R. = 2.10, $p=0.007$) or product (O.R. = 1.73, $p=0.032$) are significantly higher. 
Further, respondents find \prohibited{} ads to be unclear in their description (O.R. = 1.89, $p=0.042$).
%, perhaps because they are attempting to obscure the true, potentially prohibited, nature of the product or content they are advertising.
%
Finally, our results find evidence that participants recognize the clickbait nature of the ads we categorize as \clickbait{} (O.R. = 1.98, $p=0.003$), as well as those we categorize as more broadly \deceptive{} (O.R. = 2.46, $p=0.002$), the latter of which are likely to use attention-grabbing content to lure people to click~\cite{kanich2008spamalytics,redmiles2018examining}.

Comparatively, the odds of \opportunity{} and \healthcare{} ads being described by participants as unclear are lower than the odds of a \neutral{} (all O.R. $\leq$ 0.55, $p < 0.05$).
We also note that \opportunity{} ads, despite having higher odds of being described as irrelevant (O.R. = 1.4, $p=0.020$), have lower odds of being described as pushy than \neutral{} ads.

%\parb{Summary.} 
Overall, we find differences in both rates of dislike, and reasons for disliking across our defined ad categories. 
%Opportunity and Healthcare ads are disliked at a comparable rate to Neutral ads, and are found to be clearer in their communication by participants. 
\prohibited{}, \deceptive{}, \clickbait{}, and \sensitive{} ads 
%(which are prohibited or restricted by Facebook's policies as well) 
are found to be disliked at a higher rate than other categories, and for more severe reasons beyond irrelevance: participants recognize their clickbait-y and scammy nature; dislike the sensitive products they advertise and the advertisers selling those products; and find them unclear, potentially due to advertisers evading platform prohibitions. % loud sensationalist nature, trust them less, find them unclear, and in some instances directly dislike the advertiser or the product. We argue that exposure to such ads constitutes a problematic online experience, because participants both dislike and mistrust the content. 
As such, for the remainder of this paper we refer to the collection of these four ad categories as \problematic{}.

\subsection{How are \problematic{} ads distributed?}
\label{sec:distribution_analysis}
%Next, we examine how these problematic ads are distributed across participants, and aim to identify any skews in that distribution (RQ2).  We first examine skews for each ad category before examining each participant's {\em ad diet}.

%First, we measure skew at the level of each ad category, to understand whether it has a skewed prevalence across our panel. Second, we investigate skews at the level of each participant's {\it ad diet}, to measure if some participants have increased exposure to problematic ads in their individual experience, compared to the panel.

%\subsubsection{Skewed distribution of Problematic ads}
% ===== gini coefficient vals on counts ====
% Benign 0.48
% Opportunity 0.591
% Healthcare 0.597
% Deceptive 0.695
% Clickbait 0.655
% Potentially Prohibited 0.672
% Sensitive 0.622
% Financial 0.65

% ==== 0.8 on x-axis for CDF ====
% Neutral: x-axis @ 80%: 60
% Healthcare: x-axis @ 80%: 47
% Opportunity: x-axis @ 80%: 48
% Sensitive: x-axis @ 80%: 45
% Financial: x-axis @ 80%: 42
% Clickbait: x-axis @ 80%: 39
% Potentially Prohibited: x-axis @ 80%: 38
% Deceptive: x-axis @ 80%: 36

\begin{figure}[t]
\centering
  \centering
  \includegraphics[width=1\linewidth]{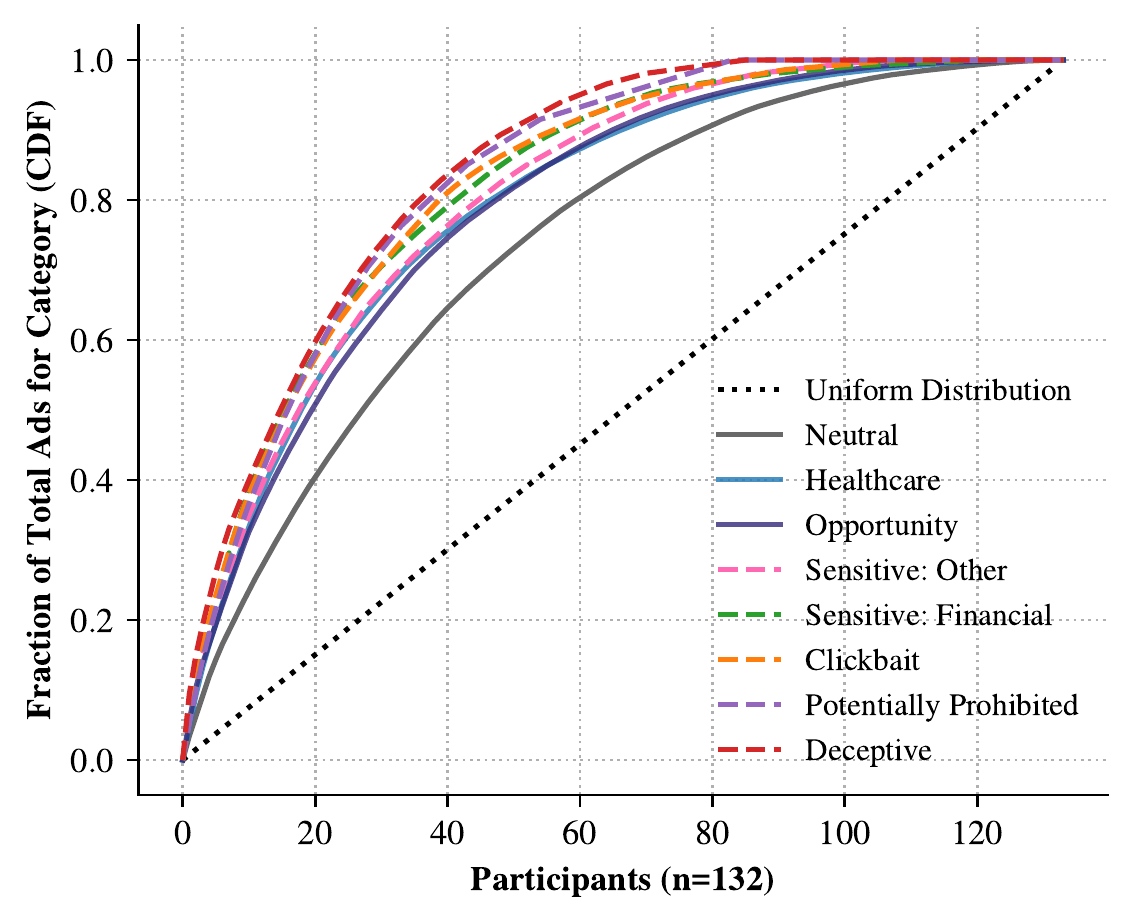}
\caption{Cumulative Distribution Function (CDF) of impressions, showing what fraction of each ad category's total ($y$-axis) is contributed by how many participants ($x$-axis), given 132 total active participants. %Non-problematic and Neutral ads are shown as solid lines, problematic ad categories shown as dashed lines. Diagonal dotted line of uniform distribution also shown, which represents hypothetical scenario where every participant has equal ad counts for an ad category.
}
\label{fig:distribution_cdf}
\end{figure}

To understand how each ad category is distributed over our panel, we investigate the skew in its distribution over our participants:
% what fraction of the category in our data is contributed by how many participants
Figure~\ref{fig:distribution_cdf} shows a cumulative distribution function (CDF) for all ads in each category. %is a useful tool to measure this effect. 
We also employ the Gini coefficient
% (defined in Section~\ref{sec:meth:analysis})
to precisely quantify this inequality. %; the Gini coefficient  is a popular metric in economics to  measure inequality among values of a frequency distribution. 
%Figure~\ref{fig:distribution_cdf} shows a CDF for each category, across our participants. 
While highly recurrent impressions of ads are relatively sparse in our data---a median of two impressions per ad per participant---we account for the frequency of impressions
% with which each ad was shown to a participant
in this analysis as well.
% Therefore the CDF shows fractions of ad {\it impressions}, when an ad was shown multiple times to a participant.
%

First, we observe that \neutral{} ads are not uniformly distributed, as observed by the distance from a uniform distribution.
Because of this inherent skew in ad distribution, we treat \neutral{} (Gini = 0.48) as the baseline for comparison.
Second, we see that \healthcare{} (Gini = 0.60) and \opportunity{} (Gini = 0.59) ads are more skewed (i.e., less uniformly distributed) than \neutral{}. This may be because \healthcare{} and \opportunity{} ads focus on narrower themes, and may be more personalized to users by advertisers or the platform.
Third, we find that all five \problematic{} categories are more skewed across participants than \neutral{}. In these  categories, we note the following order from least to most skewed: \sensitiveother{} (Gini = 0.62), \sensitivefinancial{} (0.65), \clickbait{} (0.66), \prohibited{} (0.67), and \deceptive{} (0.69).
% \elissa{I actually have no idea if people do stats testing on Gini coefficients, but someone is likely to ask if 0.60 is really that different from 0.62 so you might want to check or consider phrasing a little more gently because even if statistically significant some of the effect sizes are arguably small in comparison with Health/Opportunity -- but still big compared with neutral}
% \ali{backspaced they're more skewed than non-problematic, only even more skewed compared to neutral}
%
To offer a concrete example of this skew: 80\% of the \deceptive{} ad impressions (0.8 on $y$-axis) are delivered to just 36 participants ($x$-axis), compared to \healthcare{}, where the same fraction of impressions are delivered to 47 participants (or 60 participants in the case of \neutral{}).

%\elissa{this paragraph felt confusing here without the next set of data, maybe move to the end of this section?} This suggests that unlike Neutral ads, e.g. for promotions and local events, fewer participants in our panel come across problematic advertising. However, at least \todo{X\%} of the ads any participant sees are $problematic$. Further, the identified skew means that a small subset of our panel experience problematic advertising at a much higher frequency than others.

\begin{figure}[t]
\centering
  \centering
  \includegraphics[width=1\linewidth]{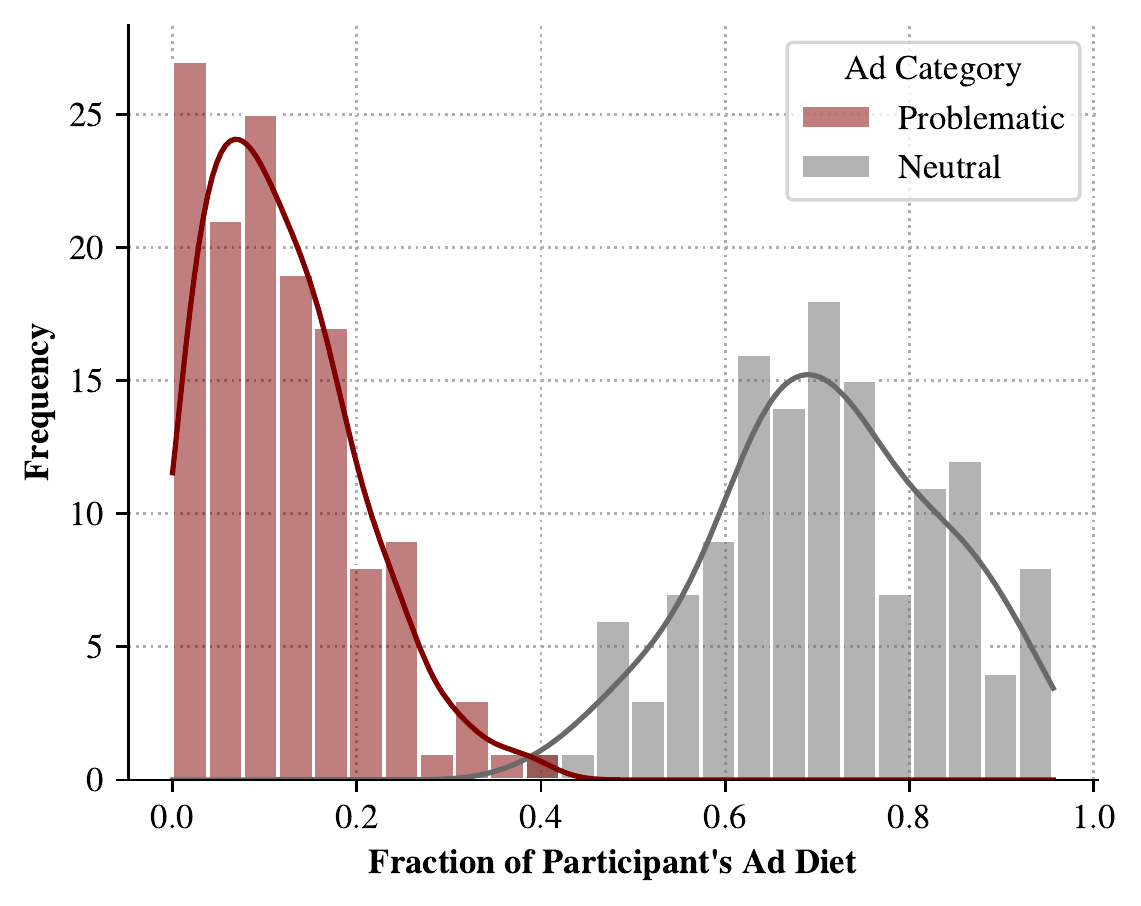}
\caption{Fractions of exposure to \neutral{} and \problematic{} ads, out of participants' overall ad diet. We factor in frequency of seeing an ad while computing fractions. Smoothed lines are kernel density estimates (KDE) of the probability distribution.}
\label{fig:exposure_distribution}
\end{figure}

% \subsubsection{Skewed individual exposure to \problematic{} ads}

Next, we focus on how individual-level exposure to \problematic{} ads vary for our participants.
%
%While we find high-level skews in the distribution of such ads compared to \neutral{}, 
First, we note that data contributions themselves are inherently skewed, since participants have varying rates of Facebook use.
To control for these differences, we look at the fraction of every participant's {\em ad diet}, i.e., all ads seen by them during the study, that consisted of \neutral{} vs. \problematic{} categories.
Figure~\ref{fig:exposure_distribution} shows the frequency distribution of these fractions across our panel.

We first observe that on average, a higher fraction of our panel's ad diet is composed of \neutral{} ads ($\mu = 0.71$, $\sigma=0.12$), compared to \problematic{} ($\mu = 0.12$, $\sigma=0.08$).
%or %non-problematic \elissa{if reverting to the old figure remove this note b/c it feels a little weird? instead I would put problematic in special font}
%Healthcare and Opportunity ads ($\mu=0.17$, $\sigma=0.09$).
%
Confirming our findings in the prior section, the distribution of \problematic{} has a heavier tail, suggesting that certain participants in our panel have increased exposure to these ads compared to the average.
%However, we note here that the distribution of fraction problematic has a heavier tail
%
This observation is supported by measuring the skewness of these distributions, a statistical measure of asymmetry of a probability distribution. Recall that positive skew implies a distribution has a long right tail, while a negative skew means the left tail is longer.
We measure the skewness for \neutral{} in Figure \ref{fig:exposure_distribution}\linebreak as -0.11, and for \problematic{} as 0.84.
%Similarly, the skewness of fraction non-problematic (blue) is 0.83. 
%
These differences imply that despite the average exposure to \neutral{} ads in our panel being 71\%, certain participants exist at the long left tail of this distribution, who are shown fewer \neutral{} ads, and a higher fraction of \problematic{} ads.

% ========== fig:exposure_distribution stats ===============
% --- medians ---
% problematic: 0.10
% neutral: 0.71
% non-problematic: 0.15
% --- means ---
% problematic: 0.12
% neutral: 0.71
% non-problematic: 0.17
% --- skewness ---
% problematic: 0.837
% neutral: -0.111
% non-problematic: 0.834
% --- std ---
% problematic: 0.081
% neutral: 0.123
% non-problematic: 0.093
\input{tables/demographic_skew.tex}

We next examine these participants who are shown 
%\elissa{consume implies some level of control on the users' part when really they don't have control over this 
a higher fraction of \problematic{} ads. Specifically, we investigate whether for any particular demographic groups, the \problematic{} ads constitute a higher fraction of ad diets.
Table~\ref{tab:demographic_skew} shows coefficients of six linear models that we build to examine the relationship between participant demographics and fraction of \problematic{} ads among the ads they encountered.
%
%For each ad category, we model the fraction that this category constitutes in the participants overall diets (a real number) as a dependent variable, and participants' self-described demographic attributes (Boolean) as independent variables.
%
The intercept shows the average fraction in the ad diets of participants for whom all independent demographic variables are \texttt{false}, i.e., white, non-Hispanic men, born in 1980 or after, without a college degree.
The proportion of these participants' ad diets that is composed of \problematic{} ads is 12\% (first column in Table~\ref{tab:demographic_skew}).
All statistically significant coefficients in the table mark biases in comparison to that baseline.

% === nice examples from Piotr ===
% For example, women's ad diets are composed of 6.4\% points (95\% CI: 4-9\%) fewer \problematic{} ads than those who do not identify as women ($\textrm{12\%} + (-\textrm{6.4\%}) = \textrm{5.6\%}$).
% for older: ($\textrm{12\%} + \textrm{5.1\%} = \textrm{17.1\%}$)

We find that the ad diets of older participants, born before 1980, are (additively) composed of 5.1\%  more \problematic{} ads (CI: 2-8\%) than younger participants.
Women's ad diets are composed of 6.4\% fewer \problematic{} ads (CI: 4-9\%) than those who do not identify as women---largely because women see 4.5\% fewer \sensitivefinancial{} ads (CI: 2-7\%).
We also note that older participants' ad diets are composed of higher fractions of \deceptive{} (1.1\%, CI: 0-2\%), and \clickbait{} ads (1.3\%, CI: 1-3\%). 
%
%\sensitivefinancial ads constitute a 4.5\% smaller fraction (95\% CI: 2-7\%) of ads seen by women.
%
Ad diets of Black participants contain 1.3\% (CI: 0-2\%) more \clickbait{} ads than those of white or Asian participants in our panel.
However, older participants and Hispanic participants ad diets have slightly lower
% odds of containing a high
fraction of \prohibited{} ads, 0.3\% (CI: 0-1\%) and 0.7\% (CI: 0-1\%) respectively, potentially because these ads target products assumed by advertisers or the platforms not to be of interest to these groups.
To account for possible variance in participants' privacy behavior (e.g. changing ad preferences), we model their awareness of privacy settings as an additional independent variable in Table~\ref{tab:demographic_skews_extra}.
We find that privacy awareness does not have any significant effect on the disparate exposure that we observe, and demographic skews similar to those in Table~\ref{tab:demographic_skew} persist.
Demographic skews for other ad categories are also shown in Table~\ref{tab:demographic_skews_extra}.
%
%In summary, we find that problematic ads skew towards older participants in our panel, and those who identify as men.

%This provides more insight into the demographic composition of the participants who are exposed to clickbait, scam, and possibly harmful ads in our panel. 

\begin{figure}[t!]
  \centering
  \includegraphics[width=1\linewidth]{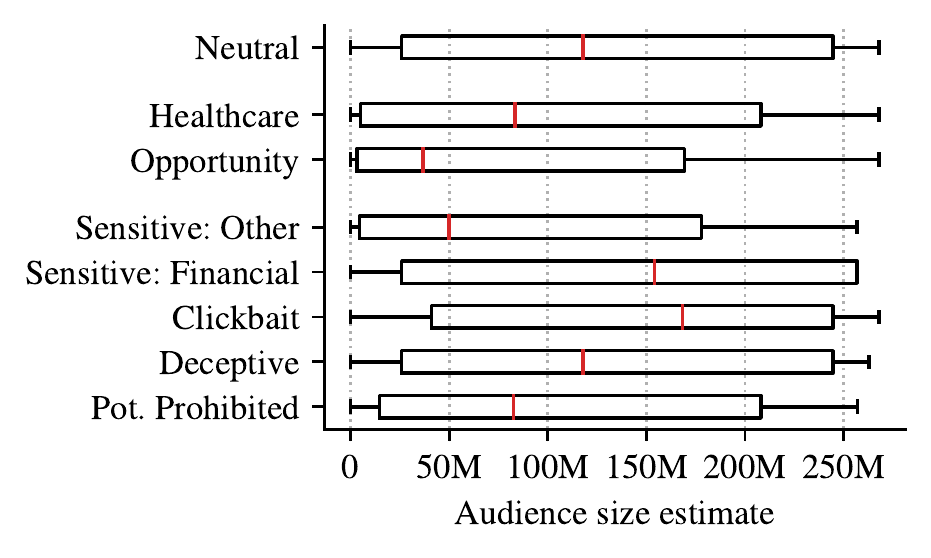}
\caption{Audience size distributions of different ad categories. The red vertical lines mark the median audience size, the box indicates the 25th and 75th percentile, and the whiskers extend from the box by 1.5x of the inter-quartile range (IQR).} 
\label{fig:audience_sizes}
\end{figure}

\input{sec53.tex}

%% file: tables/dislike_reasons_regression.tex
% TODO: switch all to \scriptsize{}

\begin{table*}[ht]
\centering
\rowcolors{4}{white}{gray!25}
\resizebox{.9\textwidth}{!}{
\begin{tabular}{lccccc|c|cc}
  {\bf Dislike Reason} & \multicolumn{8}{c}{\shortstack{Odds Ratio \\ \scriptsize{[95\% CI]}}} \\
  \toprule
  & \makecell{\bf Pot. \\ \bf Prohibited} & {\bf Deceptive} & {\bf Clickbait} & \makecell{\bf Sensitive:\\ \bf Financial} & \makecell{\bf Sensitive:\\ \bf Other} & {\bf Problematic} & {\bf Opportunity} & {\bf Healthcare}\\
 %\cmidrule(lr){2-2}
  % \hline \\[-0.5ex]
  \midrule
intercept & \shortstack{0.03$^{***}$\\ \scriptsize{[0.02, 0.06]}} & \shortstack{0.036$^{***}$\\ \scriptsize{[0.02, 0.07]}} & \shortstack{0.103$^{***}$\\ \scriptsize{[0.07, 0.16]}} & \shortstack{0.16$^{***}$\\ \scriptsize{[0.11, 0.24]}} & \shortstack{0.052$^{***}$\\ \scriptsize{[0.03, 0.09]}} & \shortstack{0.403$^{***}$\\ \scriptsize{[0.29, 0.55]}} & \shortstack{0.236$^{***}$\\ \scriptsize{[0.17, 0.32]}} & \shortstack{0.231$^{***}$\\ \scriptsize{[0.17, 0.32]}} \\ 
{\bf advertiser} & \shortstack{0.438\\ \scriptsize{[0.16, 1.21]}} & \shortstack{0.701\\ \scriptsize{[0.36, 1.37]}} & \shortstack{0.937\\ \scriptsize{[0.55, 1.6]}} & \shortstack{0.679\\ \scriptsize{[0.4, 1.14]}} & \shortstack{\textbf{2.101}$^{**}$\\ \scriptsize{[1.22, 3.63]}} & \shortstack{0.99\\ \scriptsize{[0.71, 1.39]}} & \shortstack{1.437\\ \scriptsize{[0.95, 2.17]}} & \shortstack{1.098\\ \scriptsize{[0.69, 1.76]}} \vspace{.2em}\\ 
{\bf clickbait} & \shortstack{0.657\\ \scriptsize{[0.27, 1.58]}} & \shortstack{\textbf{2.465}$^{**}$\\ \scriptsize{[1.37, 4.43]}} & \shortstack{\textbf{1.983}$^{**}$\\ \scriptsize{[1.26, 3.13]}} & \shortstack{0.93\\ \scriptsize{[0.58, 1.5]}} & \shortstack{1.265\\ \scriptsize{[0.69, 2.32]}} & \shortstack{\textbf{1.472}$^*$\\ \scriptsize{[1.07, 2.03]}} & \shortstack{1.124\\ \scriptsize{[0.74, 1.7]}} & \shortstack{0.721\\ \scriptsize{[0.44, 1.18]}} \vspace{.2em}\\ 
{\bf design} & \shortstack{0.734\\ \scriptsize{[0.36, 1.49]}} & \shortstack{1.098\\ \scriptsize{[0.58, 2.08]}} & \shortstack{0.935\\ \scriptsize{[0.58, 1.5]}} & \shortstack{0.743\\ \scriptsize{[0.48, 1.15]}} & \shortstack{1.168\\ \scriptsize{[0.67, 2.05]}} & \shortstack{0.884\\ \scriptsize{[0.65, 1.2]}} & \shortstack{1.175\\ \scriptsize{[0.81, 1.71]}} & \shortstack{1.091\\ \scriptsize{[0.73, 1.62]}} \vspace{.2em}\\ 
{\bf irrelevant} & \shortstack{1.677\\ \scriptsize{[0.99, 2.85]}} & \shortstack{1.071\\ \scriptsize{[0.68, 1.69]}} & \shortstack{1.043\\ \scriptsize{[0.74, 1.48]}} & \shortstack{\textbf{1.574}$^{**}$\\ \scriptsize{[1.15, 2.16]}} & \shortstack{1.327\\ \scriptsize{[0.87, 2.02]}} & \shortstack{\textbf{1.34}$^*$\\ \scriptsize{[1.07, 1.68]}} & \shortstack{\textbf{1.406}$^*$\\ \scriptsize{[1.05, 1.87]}} & \shortstack{1.23\\ \scriptsize{[0.91, 1.67]}} \vspace{.2em}\\ 
{\bf politicized} & \shortstack{2.754\\ \scriptsize{[0.91, 8.37]}} & \shortstack{1.759\\ \scriptsize{[0.61, 5.04]}} & \shortstack{1.359\\ \scriptsize{[0.59, 3.16]}} & \shortstack{0.78\\ \scriptsize{[0.3, 2.02]}} & \shortstack{0.15\\ \scriptsize{[0.02, 1.27]}} & \shortstack{1.128\\ \scriptsize{[0.62, 2.05]}} & \shortstack{0.818\\ \scriptsize{[0.38, 1.75]}} & \shortstack{1.797\\ \scriptsize{[0.85, 3.78]}} \vspace{.2em}\\ 
{\bf product} & \shortstack{0.832\\ \scriptsize{[0.41, 1.7]}} & \shortstack{0.954\\ \scriptsize{[0.56, 1.64]}} & \shortstack{1.078\\ \scriptsize{[0.68, 1.7]}} & \shortstack{0.997\\ \scriptsize{[0.66, 1.52]}} & \shortstack{\textbf{1.734}$^*$\\ \scriptsize{[1.05, 2.88]}} & \shortstack{1.048\\ \scriptsize{[0.78, 1.4]}} & \shortstack{0.987\\ \scriptsize{[0.67, 1.45]}} & \shortstack{0.705\\ \scriptsize{[0.45, 1.09]}} \vspace{.2em}\\ 
{\bf pushy} & \shortstack{0.747\\ \scriptsize{[0.28, 1.97]}} & \shortstack{1.209\\ \scriptsize{[0.6, 2.45]}} & \shortstack{0.682\\ \scriptsize{[0.37, 1.26]}} & \shortstack{1.367\\ \scriptsize{[0.83, 2.26]}} & \shortstack{0.499\\ \scriptsize{[0.22, 1.16]}} & \shortstack{1.008\\ \scriptsize{[0.7, 1.46]}} & \shortstack{\textbf{0.572}$^*$\\ \scriptsize{[0.34, 0.95]}} & \shortstack{1.505\\ \scriptsize{[0.96, 2.36]}} \vspace{.2em}\\ 
{\bf scam} & \shortstack{1.749\\ \scriptsize{[0.98, 3.12]}} & \shortstack{\textbf{1.972}$^{**}$\\ \scriptsize{[1.21, 3.21]}} & \shortstack{\textbf{1.473}$^*$\\ \scriptsize{[1.01, 2.14]}} & \shortstack{\textbf{1.45}$^*$\\ \scriptsize{[1.03, 2.05]}} & \shortstack{\textbf{2.078}$^{**}$\\ \scriptsize{[1.34, 3.21]}} & \shortstack{\textbf{1.643}$^{***}$\\ \scriptsize{[1.28, 2.1]}} & \shortstack{1.314\\ \scriptsize{[0.96, 1.8]}} & \shortstack{0.894\\ \scriptsize{[0.63, 1.28]}} \vspace{.2em}\\ 
{\bf unclear} & \shortstack{\textbf{1.891}$^*$\\ \scriptsize{[1.02, 3.5]}} & \shortstack{0.566\\ \scriptsize{[0.29, 1.12]}} & \shortstack{1.387\\ \scriptsize{[0.91, 2.11]}} & \shortstack{1.109\\ \scriptsize{[0.75, 1.64]}} & \shortstack{0.798\\ \scriptsize{[0.46, 1.38]}} & \shortstack{1.137\\ \scriptsize{[0.86, 1.51]}} & \shortstack{\textbf{0.55}$^{**}$\\ \scriptsize{[0.37, 0.81]}} & \shortstack{\textbf{0.592}$^*$\\ \scriptsize{[0.39, 0.9]}} \vspace{.2em}\\ 
{\bf uncomfortable} & \shortstack{1.603\\ \scriptsize{[0.59, 4.39]}} & \shortstack{0.798\\ \scriptsize{[0.33, 1.95]}} & \shortstack{1.382\\ \scriptsize{[0.71, 2.69]}} & \shortstack{0.491\\ \scriptsize{[0.22, 1.12]}} & \shortstack{0.642\\ \scriptsize{[0.23, 1.82]}} & \shortstack{0.915\\ \scriptsize{[0.56, 1.48]}} & \shortstack{1.274\\ \scriptsize{[0.72, 2.26]}} & \shortstack{0.631\\ \scriptsize{[0.3, 1.33]}} \\ 
  % 12 & sd__(Intercept) & \shortstack{1.178\\ \footnotesize{[NA, NA]}} & \shortstack{1.483\\ \footnotesize{[NA, NA]}} & \shortstack{1.101\\ \footnotesize{[NA, NA]}} & \shortstack{1.057\\ \footnotesize{[NA, NA]}} & \shortstack{1.183\\ \footnotesize{[NA, NA]}} & \shortstack{1.114\\ \footnotesize{[NA, NA]}} & \shortstack{0.719\\ \footnotesize{[NA, NA]}} & \shortstack{0.753\\ \footnotesize{[NA, NA]}} \\ 
   \midrule
   $N$ & 1152 & 1213 & 1308 & 1386 & 1227 & 2018 & 1408 & 1359\\
    %Conditional $R^2$ & 0.339 & 0.427 & 0.291 & 0.275 & 0.350 & 0.288 & 0.162 & 0.172\\
\end{tabular}
}
\caption{Odds ratios and 95\% confidence intervals for mixed-effects logistic regression models, with a random effect term for respondents. Each model examines association between ad category and dislike reasons in survey responses. Each column shows shows one model, where dependent variable is the category (boolean) in the column header. Independent variable (rows) are respondents' binary responses for different dislike reasons. Each model is fit on responses for the category in the column and Neutral ads, so odds ratios should be interpreted as comparisons with the Neutral baseline. All highly disliked categories from Figure~\ref{fig:survey_responses} are also modeled together in the ``Problematic" column. $p<0.001^{***}$; $p<0.01^{**}$, $p<0.05^*$.
%Conditional $R^2$ values, i.e. variance explained by the entire model, including both fixed and random effects~\cite{nakagawa2013general}, shown for each model.
}
\label{tab:dislike_reasons}
\end{table*}

%% file: tables/demographic_skew.tex
\begin{table*}[ht]
\centering
\renewcommand{\arraystretch}{1}
\rowcolors{2}{gray!25}{white}
\resizebox{.8\textwidth}{!}{
\begin{tabular}{lc|ccccc}
{\bf Variable} & \multicolumn{6}{c}{\shortstack{\textbf{Estimate ($\beta$)} \\ \scriptsize{[95\% CI]}}} \\
\toprule
 & {\bf Problematic} & \makecell{\bf Pot. \\\bf Prohibited} & {\bf Deceptive} & {\bf Clickbait} & \makecell{\bf Sensitive: \\\bf Financial} & \makecell{\bf Sensitive: \\\bf Other}\\
  \midrule
  Intercept & \shortstack{0.12$^{***}$\\ \scriptsize{[0.09, 0.15]}} & \shortstack{0.01$^{***}$\\ \scriptsize{[0.01, 0.01]}} & \shortstack{0.008\\ \scriptsize{[0, 0.02]}} & \shortstack{0.012\\ \scriptsize{[0, 0.02]}} & \shortstack{0.07$^{***}$\\ \scriptsize{[0.04, 0.1]}} & \shortstack{0.02$^{**}$\\ \scriptsize{[0.01, 0.03]}} \vspace{.2em}\\ 
  {\bf Gender}: Woman & \shortstack{\textbf{-0.064}$^{***}$\\ \scriptsize{[-0.09, -0.04]}} & \shortstack{-0.002\\ \scriptsize{[0, 0]}} & \shortstack{-0.005\\ \scriptsize{[-0.01, 0]}} & \shortstack{-0.008\\ \scriptsize{[-0.02, 0]}} & \shortstack{\textbf{-0.045}$^{***}$\\ \scriptsize{[-0.07, -0.02]}} & \shortstack{-0.004\\ \scriptsize{[-0.02, 0.01]}} \vspace{.2em}\\
  {\bf Race}: Black & \shortstack{0.025\\ \scriptsize{[-0.01, 0.06]}} & \shortstack{-0.001\\ \scriptsize{[0, 0]}} & \shortstack{0.006\\ \scriptsize{[0, 0.02]}} & \shortstack{\textbf{0.013}$^*$\\ \scriptsize{[0, 0.02]}} & \shortstack{0.004\\ \scriptsize{[-0.02, 0.03]}} & \shortstack{0.002\\ \scriptsize{[-0.01, 0.02]}} \vspace{.2em}\\ 
  {\bf Race}: Asian & \shortstack{-0.002\\ \scriptsize{[-0.04, 0.04]}} & \shortstack{0.001\\ \scriptsize{[0, 0.01]}} & \shortstack{-0.003\\ \scriptsize{[-0.02, 0.01]}} & \shortstack{0.005\\ \scriptsize{[-0.01, 0.02]}} & \shortstack{-0.007\\ \scriptsize{[-0.04, 0.03]}} & \shortstack{0.002\\ \scriptsize{[-0.02, 0.02]}} \vspace{.2em}\\
  {\bf Ethnicity}: Hispanic & \shortstack{0.023\\ \scriptsize{[-0.03, 0.08]}} & \shortstack{\textbf{-0.007}$^*$\\ \scriptsize{[-0.01, 0]}} & \shortstack{0.005\\ \scriptsize{[-0.01, 0.02]}} & \shortstack{-0.007\\ \scriptsize{[-0.03, 0.01]}} & \shortstack{0.036\\ \scriptsize{[-0.01, 0.08]}} & \shortstack{-0.003\\ \scriptsize{[-0.02, 0.02]}} \vspace{.2em}\\ 
  {\bf Education}: college and above & \shortstack{0.01\\ \scriptsize{[-0.02, 0.04]}} & \shortstack{-0.002\\ \scriptsize{[0, 0]}} & \shortstack{0.004\\ \scriptsize{[-0.01, 0.01]}} & \shortstack{0.01\\ \scriptsize{[0, 0.02]}} & \shortstack{-0.003\\ \scriptsize{[-0.03, 0.02]}} & \shortstack{0\\ \scriptsize{[-0.01, 0.01]}} \vspace{.2em}\\ 
  {\bf Age}: Gen-X and older & \shortstack{\textbf{0.051}$^{***}$\\ \scriptsize{[0.02, 0.08]}} & \shortstack{\textbf{-0.003}$^*$\\ \scriptsize{[-0.01, 0]}} & \shortstack{\textbf{0.011}$^*$\\ \scriptsize{[0, 0.02]}} & \shortstack{\textbf{0.017}$^{**}$\\ \scriptsize{[0.01, 0.03]}} & \shortstack{0.017\\ \scriptsize{[-0.01, 0.04]}} & \shortstack{0.009\\ \scriptsize{[0, 0.02]}} \vspace{.2em}\\ 
   %\midrule
   %$R^2$ & 0.215 & 0.086 & 0.066 & 0.143 & 0.151 & 0.021 \\
\end{tabular}
}

\caption{Coefficients of linear regression models, with 95\% confidence intervals, modeling the relationship between exposure to \problematic{} ads and participants' demographics. Dependent variable (columns): fraction of ad type, out of total ad diet. Independent variable (rows): participant demographics. Union of all problematic ad types modeled in the \problematic{} column. \newline $p<0.001^{***}$; $p<0.01^{**}$, $p<0.05^*$.}
\label{tab:demographic_skew}
\end{table*}

%% file: sec53.tex
\subsection{Who is responsible for skews?}
\label{sec:targeting_analysis}
With a better understanding of which participants have increased exposure to problematic ads, we next identify the reasons behind these differences.
As discussed in Section~\ref{sec:background}, whether a particular user sees an ad on Facebook % based on the final outcome of whether a user sees a particular ad 
is affected by two main factors:
(a) the user has to be among the audience targeted by the advertiser;
(b) Facebook's ad delivery optimization considers the ad relevant to the user, which contributes to it winning an auction~\cite{FacebookAdAuctions}.
%Facebook's selection is a partially result of their ad delivery and personalization algorithms~\cite{FacebookAdAuctions}, which we will refer to in combination as \emph{delivery optimization}.
%From our vantage point as an external-to-platform browser extension,  it is not possible to determine exactly how large a role the delivery optimization played in the delivery of all the ads that our participants see.
Thus, one can expect that when the advertiser targets a larger audience, the delivery optimization has more influence in selecting the actual audience.
With this intuition, we start by investigating audience size across our ad categories.

%Sensitive: Financial, Clickbait, and Deceptive ads tend to be at least as inclusive as Neutral ads. Potentially prohibited and Sensitive ads target much smaller audiences.}

\begin{figure*}[t!]
  \centering
  \includegraphics[width=1\linewidth]{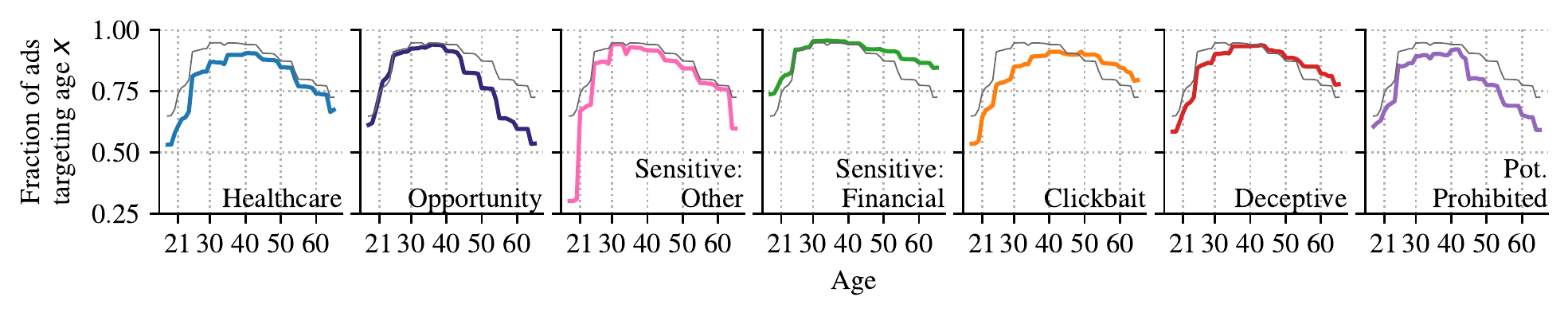}
\caption{Fraction of ads that include given age ranges in their targeting. The thin line in each panel shows the fraction among \neutral{} ads for easier comparison.}
\label{fig:targeting_ages}
\end{figure*}

% \parb{Audience Sizes.} We then investigate how much the targeting parameters used by advertisers narrow down the audiences.
As described in Section~\ref{sec:meth:analysis}, we query Facebook's APIs to obtain audience sizes for each of our collected ads---
% To this end, we pose as an advertiser and obtain audience size estimates for each combination of targeting criteria that we encounter in our dataset using an series of automated API calls.
% Note that these estimates are not accessible for the ads that use Custom Audiences - those are only known to the owners of these CAs.
Figure~\ref{fig:audience_sizes} shows the distributions of these audience sizes broken down by ad category.
Observing \problematic{} categories, we find that the median target audience sizes for \sensitivefinancial{} (153.9M) and \clickbait{} (168.2M) ads are larger than for \neutral{} ads (117.9M); a pairwise Kruskal-Wallis~\cite{kruskal1952use} test rejected the null hypothesis that the medians are equal ($p=0.001$ for both tests).
This implies that Facebook exercises more control for picking the audience subset for these categories.
On the other hand, median audience sizes for \prohibited{} (82.6M) and \sensitiveother{} (49.9M) ads
are significantly smaller than \neutral{} ($p=0.006$ and $p<0.001$, respectively), indicating that advertisers for these ads more precisely specify the audiences they want to reach.
We also note that audience sizes for \opportunity{} (36.8M) and \healthcare{} (83.4M), considered non-problematic in this study, are actually smaller than \neutral{} ($p < 0.001$).
% In fact, 24.9\% of \opportunity{} ads, our most narrowly targeted category, target audiences smaller than 3 million users, corresponding to less than 1.1\% of possible targettable users.

% \begin{figure}[t!]
%   \centering
%   \includegraphics[width=.9\linewidth]{plots/rq3_prevalence.pdf}
% \caption{Prevalence of different targeting options.}
% \label{fig:targeting_prevalence}
% \end{figure}

Next, we investigate what %Given this variance in size, we investigate what 
targeting options advertisers use to scope these various audiences.
%(Figure~\ref{fig:targeting_prevalence}).
%
We find that the most used targeting option is age: nearly half the ads use some form of age targeting (49.7\%).
Around a quarter of ads use Custom Audiences~\cite{fbcustomaudiences} (25.6\%) and platform-inferred user interests (26.9\%); 
On the other hand, advertisers for 21.2\% of the ads in our dataset don't change the targeting criteria at all, and use the default targeting of all U.S. adults (267 million users). 
Finally, we find that only 12.1\% ads in our data specifically target by gender; a vast majority use the default option of targeting all genders.
Note that these percentages do not sum up to 100\% because each ad can be targeted using multiple targeting criteria.
%
% Below, we detail how each of these targeting options are used in our data.
Below, we detail how age, custom audiences, interests and default targeting are used in our data. 

\begin{figure}[t]
  \centering
  \includegraphics[width=.9\linewidth]{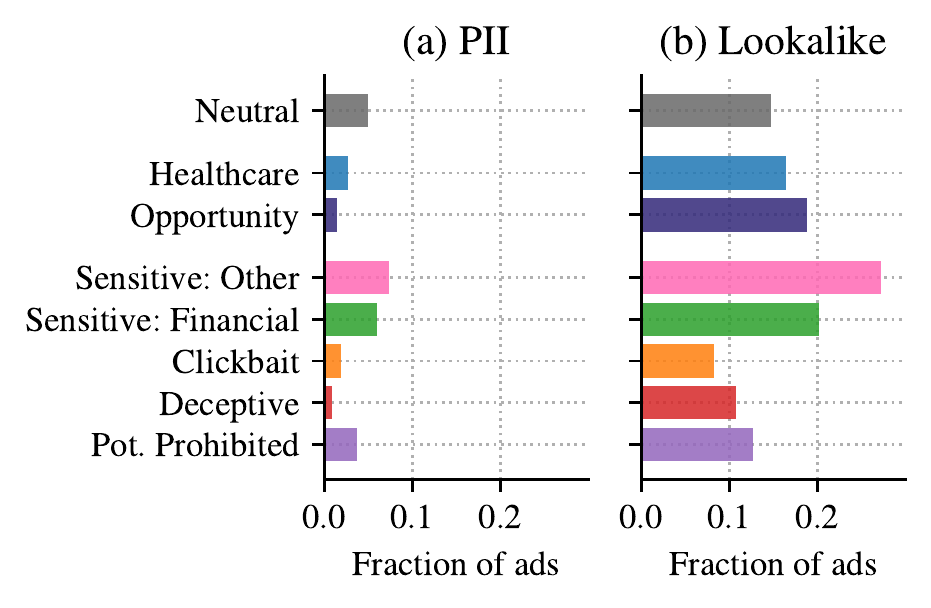}
\caption{Prevalence of two types of Custom Audiences: based on A) Personally Identifiable Information and B) Lookalike Audiences. Despite their high prevalence, Lookalike Audiences are the most opaque of targeting tools.}
\label{fig:ca_prevalence}
\end{figure}

\input{tables/default_targ_demographic_skew.tex}

\parb{Age. } 
% We observe major differences across ad categories in how advertisers use age---the most used targeting criterion in our data.
% Facebook ads target users aged 18 and over unless the advertiser chooses to narrow down this age, or extend it to include users as young as 13. 
% In our dataset only 197 ads (0.6\% of the dataset) included users below the age of 18 and 95\% of those were \neutral{}. 
% We therefore focus on ads targeting adult users.
Figure~\ref{fig:targeting_ages} shows the fraction of ads that include users of a given age in their targeting; fractions of all ages are presented together as a line, which can be perceived as a function of age.
Each panel shows this function for a different ad category, and also features the function for Neutral ads for easier comparison.
% A category-specific line above the (gray) Neutral line signifies the age ranges included a higher fraction of ads from this category than by neutral ads.
A category-specific line above the (gray) Neutral line signifies that the age group was more often targeted with ads of that category compared to \neutral{} ads.
% We note that the line for \healthcare{} ads is entirely below the \neutral{} ads, which means these ads target narrower age ranges.
% Further, a smaller fraction of \opportunity{} ads include older users than \neutral{} ads, implying that advertisers are excluding these users from opportunities. 
Focusing on \problematic{} categories, ads for \sensitiveother{} often exclude users aged 18-21. This can be explained by the prevalence of ads for alcoholic beverages in this category, selling of which to individuals below 21 is illegal in the US.
% \sensitivefinancial{} ads rely on age targeting less than any other ad category; still, a quarter of them exclude the youngest adults.
\sensitivefinancial{}, \clickbait{} and \deceptive{} ads include older audiences at a higher rate than Neutral ads, which could explain why \deceptive{} and \clickbait{} skews towards older users in our panel.
Similarly, \prohibited{} ads also exclude users over the age of 45.
These differences provide evidence that advertisers actively use the platform's age targeting features to find older users to show clickbait and scam content to.
This is notable, since prior work suggests that older users may be more susceptible to such content~\cite{munger2020null}.

% We noted previously that the target audiences of \clickbait{} and \deceptive{} ads are sized similarly to those of \neutral{} ads. 
% However, Figure~\ref{fig:targeting_ages} shows that while they might be similar in terms of size, they are not necessarily the same people: these two ad categories are more often targeted to older users and exclude younger users.

%\parb{Interests.} moved this to Appendix

\parb{Custom Audiences.}
%Now, we turn our attention to custom audiences.
We make an distinction between custom audiences where the advertiser provides Facebook with a list of particular individuals to target using their PII (e.g., phone number, email), and Lookalike Audiences~\cite{facebooklookalike} that Facebook creates by finding users similar to those that the advertiser provides.
The distinction is crucial because of the difference in control: the advertiser exercises complete control over who to include in the first group; however, they have little influence over the characteristics of the lookalike audiences.
% Nevertheless, previous work indicated that the demographic and political characteristics (including biases) of the source custom audience are replicated in the Custom Audiences~\cite{sapiezynski2022algorithms,speicher2018potential}.
%
Figure~\ref{fig:ca_prevalence} shows the prevalence of different types of custom audiences per ad category. 
We observe that lookalike audiences are used more often than PII custom audiences for all categories.
We also note that as many as a quarter of \sensitiveother{} ads were targeted using Lookalike Audiences.
This suggests that while advertisers use the platform's tool to find vulnerable audiences (e.g., Figure~\ref{fig:targeting_ages}), they often outsource this role to the platform, especially when targeting for sensitive themes like weight loss or gambling.
%
% Advertisers use Facebook to find people similar to those who engage with content relating to mental health conditions (e.g., body image disorders, alcohol addiction) more than any other kind of ad category.

\parb{Interests.}
\label{app:interest_targeting}
Precise targeting by inferred interests is one of the features that distinguishes online behavioral advertising from traditional advertising models. 
A total of 6,028 unique interests were used to target our participants, including highly specific and sensitive inferences pertaining to health (``Multiple sclerosis awareness'', ``Fibromyalgia awareness''), sexuality (``LGBT community'', ``Gay Love''), religion (``Evangelicalism'', ``Judaism''), and others.
It is perhaps surprising that a majority of ads in our dataset (73.1\%) do not actually use this functionality. 
Table~\ref{tab:targeting_interests} shows the most commonly targeted interests for each ad category.

\parb{Default Targeting. }
Finally, we investigate the delivery of ads that used the default targeting (i.e., the advertiser included all U.S. adults in their target audience).
This allow us to observe the behavior of the delivery optimization in cases where the skew can not be attributed to the advertiser's actions.
To identify skews in delivery, we run a series of linear models, shown in Table~\ref{tab:default_tar_demographic_skew}, to examine the relation between fraction of problematic ads in ad diets and participant demographics, similar to Section~\ref{sec:distribution_analysis}.
In contrast to that analysis, however, we subset our data to only include ads that have default targeting from the advertiser.
Therefore, for each participant, we model, say, the fraction of \clickbait{} they saw that had default targeting, out of all of their default-targeted ads.
Consequently, we capture purely skews that arise due to the platform's optimization, since the advertiser specified the broadest possible targeting, and Facebook had to make its judgment of a relevant audience.
%shows the coefficients for six such linear models.
Again, the first row (intercept) shows the fraction of ad diets for participants who are non-Hispanic white, younger, and without a college education; all significant coefficients mark biases in comparison to that baseline.

Table~\ref{tab:default_tar_demographic_skew} shows that (similar to Table~\ref{tab:demographic_skew}), the effect 
for older participants seeing a 7.7\% higher fraction of \problematic{} ads (CI: 2-13\%), and women seeing 5.9\% fewer of them (CI: 1-11\%), persists, even without advertiser targeting.
Specifically, older participants' ad diets (additively) contain 4.1\% (CI: 2-6\%) more \clickbait{} than the younger participants.
We also observe a novel effect of Hispanic participants seeing 2.8\% more \deceptive{} ads (CI: 1-5\%).
This implies that while their overall ad diets might not contain a significantly higher fraction of scams (Table~\ref{tab:demographic_skew})---delivery optimization independently skews these ads towards Hispanic participants.
In further nuance, the effect of women seeing fewer \problematic{} ads can be explained by their ad diets comprising of 4.6\% fewer \sensitivefinancial{} ads (CI: 0-9\%), and 0.6\% fewer \prohibited{} ads (CI: 0-1\%) compared to participants who don't identify as women.
% \elissa{need to update various summaries to include this hispanic result and write around the fact that Black is not observed here -- also that's interesting, means platform has hispanic bias while advertisers have Black bias, could be something to note in discussion}%
% These differences provide insights into delivery optimization's role in delivering \problematic{} ads. We find  both similarities to advertiser-introduced skews, such as for older participants, and differenc skews that could
These differences provide evidence that in addition to an advertiser's targeting---or regardless of it---Facebook's delivery optimization algorithms are also responsible for skewing the delivery of \problematic{} ads.

%% file: tables/default_targ_demographic_skew.tex
\begin{table*}[t!]
\centering
\renewcommand{\arraystretch}{1}
\rowcolors{2}{gray!25}{white}
\resizebox{.8\textwidth}{!}{
\begin{tabular}{lc|ccccc}
{\bf Variable} & \multicolumn{6}{c}{\shortstack{\textbf{Estimate ($\beta$)} \\ \scriptsize{[95\% CI]}}} \\
\toprule
 & {\bf Problematic} & \makecell{\bf Pot. \\\bf Prohibited} & {\bf Deceptive} & {\bf Clickbait} & \makecell{\bf Sensitive: \\\bf Financial} & \makecell{\bf Sensitive: \\\bf Other}\\
  \midrule
  Intercept & \shortstack{0.191$^{***}$\\ \scriptsize{[0.13, 0.26]}} & \shortstack{0.013$^{***}$\\ \scriptsize{[0.01, 0.02]}} & \shortstack{0.014$^*$\\ \scriptsize{[0, 0.03]}} & \shortstack{0.023\\ \scriptsize{[-0.01, 0.05]}} & \shortstack{0.133$^{***}$\\ \scriptsize{[0.08, 0.18]}} & \shortstack{0.009$^*$\\ \scriptsize{[0, 0.02]}} \vspace{.2em}\\ 
  {\bf Gender}: Woman & \shortstack{\textbf{-0.059}$^*$\\ \scriptsize{[-0.11, -0.01]}} & \shortstack{\textbf{-0.006}$^*$\\ \scriptsize{[-0.01, 0]}} & \shortstack{-0.007\\ \scriptsize{[-0.02, 0]}} & \shortstack{-0.003\\ \scriptsize{[-0.03, 0.02]}} & \shortstack{\textbf{-0.046}$^*$\\ \scriptsize{[-0.09, 0]}} & \shortstack{0.004\\ \scriptsize{[0, 0.01]}} \vspace{.2em}\\
  {\bf Race}: Black & \shortstack{0.01\\ \scriptsize{[-0.05, 0.07]}} & \shortstack{0.002\\ \scriptsize{[0, 0.01]}} & \shortstack{0.007\\ \scriptsize{[-0.01, 0.02]}} & \shortstack{0.011\\ \scriptsize{[-0.02, 0.04]}} & \shortstack{-0.007\\ \scriptsize{[-0.06, 0.04]}} & \shortstack{-0.003\\ \scriptsize{[-0.01, 0]}} \vspace{.2em}\\
  {\bf race}: Asian & \shortstack{-0.019\\ \scriptsize{[-0.1, 0.06]}} & \shortstack{-0.005\\ \scriptsize{[-0.01, 0]}} & \shortstack{-0.003\\ \scriptsize{[-0.02, 0.01]}} & \shortstack{-0.007\\ \scriptsize{[-0.04, 0.03]}} & \shortstack{-0.003\\ \scriptsize{[-0.07, 0.06]}} & \shortstack{0\\ \scriptsize{[-0.01, 0.01]}} \vspace{.2em}\\ 
  {\bf Ethnicity}: Hispanic & \shortstack{0.017\\ \scriptsize{[-0.08, 0.12]}} & \shortstack{-0.009\\ \scriptsize{[-0.02, 0]}} & \shortstack{\textbf{0.028}$^{**}$\\ \scriptsize{[0.01, 0.05]}} & \shortstack{-0.021\\ \scriptsize{[-0.06, 0.02]}} & \shortstack{0.027\\ \scriptsize{[-0.05, 0.11]}} & \shortstack{-0.008\\ \scriptsize{[-0.02, 0]}} \vspace{.2em}\\
  {\bf Education}: college and above & \shortstack{-0.033\\ \scriptsize{[-0.09, 0.02]}} & \shortstack{-0.002\\ \scriptsize{[-0.01, 0]}} & \shortstack{0\\ \scriptsize{[-0.01, 0.01]}} & \shortstack{0.005\\ \scriptsize{[-0.02, 0.03]}} & \shortstack{-0.036\\ \scriptsize{[-0.08, 0.01]}} & \shortstack{-0.001\\ \scriptsize{[-0.01, 0.01]}} \vspace{.2em}\\ 
  {\bf Age}: Gen-X and older & \shortstack{\textbf{0.077}$^{**}$\\ \scriptsize{[0.02, 0.13]}} & \shortstack{-0.003\\ \scriptsize{[-0.01, 0]}} & \shortstack{0.011\\ \scriptsize{[0, 0.02]}} & \shortstack{\textbf{0.041}$^{**}$\\ \scriptsize{[0.02, 0.06]}} & \shortstack{0.034\\ \scriptsize{[-0.01, 0.08]}} & \shortstack{-0.005\\ \scriptsize{[-0.01, 0]}} \vspace{.2em}\\ 
   %\midrule
   %$R^2$ & 0.100 & 0.095 & 0.091 & 0.119 & 0.074 & 0.040\\
\end{tabular}
}
\caption{Coefficients of linear regression models, with 95\% confidence intervals, modeling relationship between exposure to problematic ads {\it due to platform optimization}, and participants' demographics. Dependent variable (columns): fraction of category, out of total ad diet of ads with default/no advertiser targeting. Independent variable (rows): participant demographics.
% Union of all problematic ad types modeled in the Problematic column.
$p<0.001^{***}$; $p<0.01^{**}$, $p<0.05^*$.}
\label{tab:default_tar_demographic_skew}
\end{table*}

%% file: discussion.tex
%!TEX root = main.tex
\section{Concluding Discussion}\label{sec:discussion}
Our study presents three main contributions. 
%In this study, we collect the ads viewed by a diverse panel of Facebook users to understand what ads these users find problematic, how those ads were distributed across the panel, and what was the cause of the observed skews.
%While prior work has explored how ads can be delivered in a biased manner~\cite{ali2019discrimination,ali-2019-arbiters}, we build on new evidence that ads carry problematic content as perceived by users~\cite{zeng2020bad}
%\ali{very redundant sentences, CUT}
{\it First}, gathering insights from a diverse group of Facebook users, we identify a collection of \problematic{} categories of ads that were significantly more disliked, and determine participants' reasons for disliking these ads---they often mistrust these ads and recognize their deceptive nature.
{\it Second}, we observe that while these ads make up a small fraction (12\% on average) of our participants' ad diets, a subset of our panel are disproportionately exposed to them.  
%
%Specifically, older adults see a higher fraction of clickbait and scam ads, Black participants see a higher fraction of scam ads, and men see more financial ads.
% The skew in distribution of these ads is particularly concerning as it suggests that certain users have significantly worse ad diets on Facebook. %, composed of ads that they themselves dislike.
% The skew in distribution of these ads is particularly concerning as it suggests both that the distribution of opporunity ads is non-uniform, and that certain users have significantly worse ad diets on Facebook.
%
{\it Third}, using a combination of techniques, we demonstrate that some of these skews in ad distribution %are due to advertisers' targeting choices, but that the skews often 
persist without targeting from advertisers, implying that the platform's algorithms are responsible for at least some of the skews we observe.
%
% content is nuanced: finance, clickbait is personalized
% don't go too much into finance

%To our knowledge, ours is the first study to rigorously study users' lived experiences with advertising, elicit a robust understanding of problematic ads, and measure disparities in their distribution.
%
While our observations are limited to our panel, %and cannot be generalized for all Facebook users, 
our study validates anecdotal evidence~\cite{quartz_metalscom,buzzfeed_adsinc} that clickbait and scam advertising is shown to older users more often.
We show that these differences exist both due to advertisers' targeting and due to the platform's delivery optimization---which together may create a feedback loop~\cite{leqiengineering}.
We also identify instances where the overall outcomes are different than delivery optimization's biases: Black participants see a higher fraction of \clickbait{} ads (Table~\ref{tab:demographic_skew}), but only when targeted by advertisers. %when controlling for targeting, the differences are not statistically significant.
On the other hand, Hispanic participants have higher exposure to \deceptive{} ads (Table~\ref{tab:default_tar_demographic_skew}), but only within ads that are essentially untargetted by advertisers, suggesting this effect is due to the ad platform.%most broadly targeted. 
%
%This suggests that Facebook's ad platform plays an active role in determining the skew for problematic ads, in cases similar to the advertisers, and in other cases, differently than them.
%

% financial results: 
Further, we find that financial ads are shown more often to participants who identify as men, both as a system-level outcome, and when controlling for ad targeting.
As annotators, we observe that \sensitivefinancial{} ads are quite diverse---ranging from problematic offers like high APR loans to possibly useful financial tools such as savings accounts.
Thus, men in our panel are exposed to problematic financial products, as well as financial opportunities, more often.

Finally, our analysis of targeting practices shows that advertisers often cede control to the platform's optimizations -- as evidenced by the popular use of lookalike audiences (Figure~\ref{fig:ca_prevalence}) and the low usage of targeting interests (Table~\ref{tab:targeting_interests}).
This implies that advertisers are aware of the usefulness of the platform's personalization, and malicious actors could rely on these capabilities to target \problematic{} advertising.

Taken together, our results offer concrete insights into user experiences with problematic advertising and raise questions about the power of platforms in delivering these ads to users.
%, and highlight platforms' responsibilities in protecting their users from such content.

%\smallskip
%We now discuss limitations of our work, and conclude with recommendations for platforms.

\parb{Limitations.}
Our ad categories were created through pilot data collection and backed by review of platform policies and literature, including work that also examined user sentiments towards problematic advertising~\cite{zeng2020bad}. Still, categorizing ads into just seven categories diminishes some nuance within groups.
We analyze a subset of our total collected ads that we were able to annotate manually (one-third of our overall collected data); therefore, we are not able to provide insight into the complete ad diets of our participants. To minimize any selection biases in our analyzed subset, we randomly sampled ads from participants each month for annotating and surveying, but recognize important data could be missed by not assessing the complete ad diets of participants.
%
%\elissa{inclined to cut this one, esp. since we aren't rebutting it} Our participants were recruited on a rolling basis in order to maximize demographic diversity for hard-to-reach groups, which may subject our data to time-related effects like seasonal differences in advertising.
% %\elissa{ideally we would look to see if we observe time effects by putting a time variable in our models and state here that we don't see any such effects}.

% Future work may choose to dissect certain categories---for instance, examining sentiments towards different types of Opportunity ads---to more firmly understand their place in the online ad ecosystem. Platforms must also flesh out these categories to more effectively moderate against diverse types of problematic ads.

Further, our observations are only about participants' desktop browsing experiences. While we suspect that similar ads would be present on the mobile Facebook app due to the diversity of Facebook's ad placement options, we do not have direct access to that data.
We also do not have access to budgets of the ads that we observe, and therefore are not able to disambiguate whether certain advertisers are simply paying more money to Facebook, resulting in skews.
However, to control for these differences, we compare fractions of ad categories out of the ad diets that we observe for each participant (e.g., in Section~\ref{sec:distribution_analysis}).
This ensures that we compare only within participants' desktop experiences, and in the same budget-class of advertisers that were reaching them.

Additionally, we do not have access to participants' complete ad preferences, and the frequency with which they change these settings.
This limits our ability to control for participant actions such as removing ads from an advertiser, or removing a specific interest.
Prior work estimates that 10-19\% of users tweak their ad settings~\cite{habib2022identifying, im2023less}, either from the ad preferences page or from the contextual menu next to ads.
We attempt to account for such variance by factoring participants' awareness of privacy settings in Table~\ref{tab:demographic_skews_extra}, and find that disparate exposure to \problematic{} ads for older and minority participants persists.

Finally, our work currently does not provide insight on advertising's contextual harms~\cite{milano2021epistemic}; for instance, while we take an interest in sensitive ads with subject matters like gambling, we do not investigate their distribution among those with gambling addictions. 
Rather, we try to find commonalities in our panel's opinions through mixed-effects regression models, and then build our analysis on top of that data.
We leave further exploration of contextually problematic ads, such as Gak et al.~\cite{gak2022distressing}, to future work.

% While we form our codebook based on categories of ads and forms of harm previously explored in literature, and achieve sufficient inter-coder agreement during our qualitative coding process, identifying harmful content remains relatively subjective and it is possible we do not capture the full spectrum of problematic ads. We encourage future work examining the categories of problematic ads that we identify more granularly.

% Finally, our sample of 132 participants is limited to the U.S. Facebook users. Thus, our work may fail to capture categories of ads -- or biases in the distribution of problematic ads -- relevant to people from other regions. We encourage future follow-up work expanding on our analysis.

% \medskip
\parb{Recommendations.} To limit users' exposure to problematic ads, we propose changes on two levels.
{\it First}, we advocate for a more fine-grained and user-informed understanding of problematic ads, and other broader harms of advertising~\cite{ali2021measuring}.
% Our work builds on the understanding of why users dislike problematic advertising.
Currently, platforms recognize ads such as \deceptive{}, \clickbait{} and \prohibited{} as problematic, and typically include language scrutinizing them in their advertising guidelines~\cite{fb_unacceptable_business, fbaddemote}.
However, sensitive ads that present harms for users with addictions or other mental illness are less moderated.
Yet, they are still widely disliked across our diverse set of participants.
We advocate for a more refined understanding of ads with sensitive themes, and more scrutiny and moderation from platforms towards these ads.
For a more nuanced understanding of problematic ads, our work, along with \cite{zeng2020bad} and \cite{gak2022distressing} provide a start.
%A more robust understanding of this class of ads can also lead to better detection via automated moderation systems.
%Failing to appropriately moderate such problematic ads may be causing an overall dislike of online advertising, as evidenced by dislike towards \sensitivefinancial{} in our panel.
%
% the body of work detailing negative perceptions and potential harms of \deceptive{}, \clickbait{}, and \prohibited{} ads is robust---platforms, too, recognize these ads as problematic and typically include language scrutinizing them in their advertising guidelines.
% Our work introduces a novel ad category, \sensitive{}, and demonstrates that it is often equally disliked for similar reasoning as other well-backed problematic ads.
% These ads present potential harms for users who may have certain sensitivities and afflictions, like addictions or other mental health issues.
% Yet, they are still widely disliked across our diverse set of participants. We advocate for more scrutiny and moderation from platforms towards Sensitive ads; failing to appropriately moderate problematic advertising may be causing an overall dislike of online advertising, as evidenced by dislike towards ad categories like Opportunity and Healthcare, and more negative online experiences for users overall.

{\it Second}, we argue for more controls not just on moderation, but on optimization as well.
Our results demonstrate that once problematic ads circumvent a platform's review process, the platform then optimizes them towards users similar to other personalized content (e.g. Figure~\ref{fig:exposure_distribution_extra}).
%<<<<<<< HEAD
To avoid this systematic personalizing of problematic ads, platforms need policies on their delivery optimization in addition to their policies on content moderation.
% possess: (a) a detailed understanding of what kind of content is harming their users, including nuanced and gray-areas of problematic ads (our work along with \cite{zeng2020bad} and \cite{gak2022distressing} provide a start); (b) implementing
This would require platforms to constrain the optimization of problematic content for users. For instance, Facebook currently states that it demotes clickbait in content ranking~\cite{fbaddemote}, yet a demotion does not stop such content from inevitably reaching and harming some users.
% yet we find that some users still receive an outsize amount of such content, despite the findings of our work and other works suggesting % Our work supports prior findings 
% that people dislike clickbait advertising~\cite{gomez2020fail, zeng2020bad}.
There is perhaps a need for an ``optimization vacuum'' so that problematic content, even after evading moderation, cannot reach users.

% Additionally, users share negative sentiments on clickbait advertising, but mixed reviews on clickbait in general~\cite{pengnate2021effects,fb1}.
%On the user end, we aren't aware of the mechanisms or extent of demotion; and on the platform's end, our study makes it clear that \clickbait{} ads are still able to create negative online experiences for a specific subset of users. 
%=======
%To avoid this systematic personalizing of problematic ads, platforms need policies on their delivery optimization in addition to their policies on content. This would require platforms to possess: (a) a detailed understanding of what kind of content is harming their users, including nuanced and gray-areas of problematic ads (our work along with \cite{zeng2020bad} and \cite{gak2022distressing} provide a start); (b) implementing constraints on the optimization, in particular, of such content for users.
% For instance, Facebook currently states that it ``demotes" clickbait in their content ranking~\cite{fbaddemote}. On the user end, we aren't aware of the mechanisms or extent of demotion; and on the platform's end, our study makes it clear that \clickbait{} ads are still able to create negative online experiences for a specific subset of users.
%
%>>>>>>> 80c031e078ce81690780c256281771051cb833a1

We advocate for platforms to take emerging works on user experiences with problematic ads into account, and for a more urgent call for platforms to not only moderate the content users see, but also have mechanisms to suppress the delivery of problematic content, instead of optimizing for it.

% \elissa{any solutions that we want to talk about? security people are always very ``what do I do about it'' - esp. since we bring up legislative proposals. Some ideas:
% \begin{itemize}
%     \item future work to auto classify ads into categories -- arguably platform already does some of this for some of these categories, our dataset + Erics could be used as training data 
% \item using that classification, platforms could have transparent metrics about the composition of people's ad diets?
% \item We had talked about maybe restricting whether you can target by category?
% \end{itemize}
% }

% \begin{itemize}
    % \item this isn't about ``well what if the users find it relevant?" Our participants recognize scams and clickbait content when they see it
    %\item currently there is Stronger stance against clickbait, deceptive, problematic but people feel similarly about sensitive, which is less strongly scrutinized --- that should change
    % \item There should policy on the optimization AND the content
% \end{itemize}

%% file: arxiv_appendix.tex
%!TEX root = main.tex
\appendix
\section{Appendix}
\label{appendix}

\setcounter{figure}{0}
\renewcommand\thefigure{A\arabic{figure}} 
\setcounter{table}{0}
\renewcommand\thetable{A\arabic{table}} 
% \xpretocmd{\table}{\setcounter{table}{0}}{}{}
\setcounter{section}{0}
\renewcommand\thetable{A\arabic{section}} 
% TODO: reset counter for table*

%\vfill\eject

\begin{figure}[b!]
\centering
  \includegraphics[width=0.8\columnwidth]{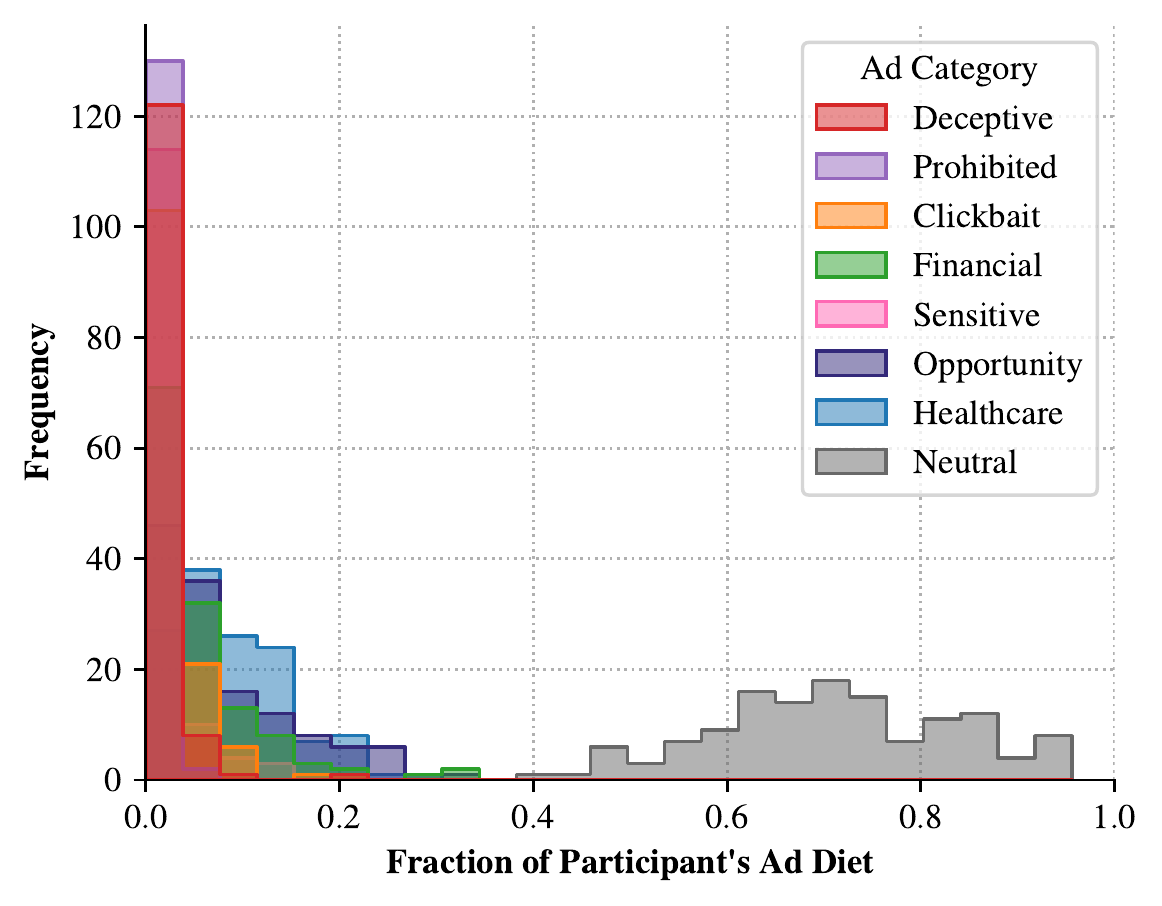}
\caption{Fractions of exposure to different ad categories, out of participants' overall ad diet. \problematic{} ads are personalized similar to others in our codebook.}
\label{fig:exposure_distribution_extra}
\end{figure}

\input{tables/joined_table.tex}

% %\section{Demographic Skews: Accounting for\\ Privacy Behavior}
% \input{tables/demographic_skew_w_privacy.tex}

% %\section{Demographic Skews:\\ Healthcare and Opportunity}
% \input{tables/healthcare_opportunity_demographic_skews.tex}

%\vfill\eject
%\section{Interest Targeting}
\input{tables/interest_targeting_condensed.tex}

\begin{figure*}[h!]
    \vspace{-50pt}
    % \subsection*{D. Example Ads}
    \section*{Example Ads}
     \centering          
     \begin{subfigure}[t]{0.3\textwidth}
         \centering
         \includegraphics[width=\textwidth]{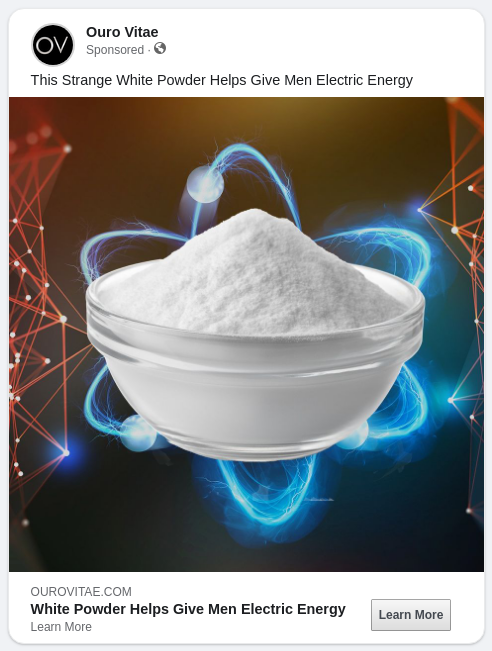}
         \caption{\prohibited{}}
         \label{fig:ad_prohibited}
     \end{subfigure}
     \hfill
     \begin{subfigure}[t]{0.3\textwidth}
         \centering
         \includegraphics[width=\textwidth]{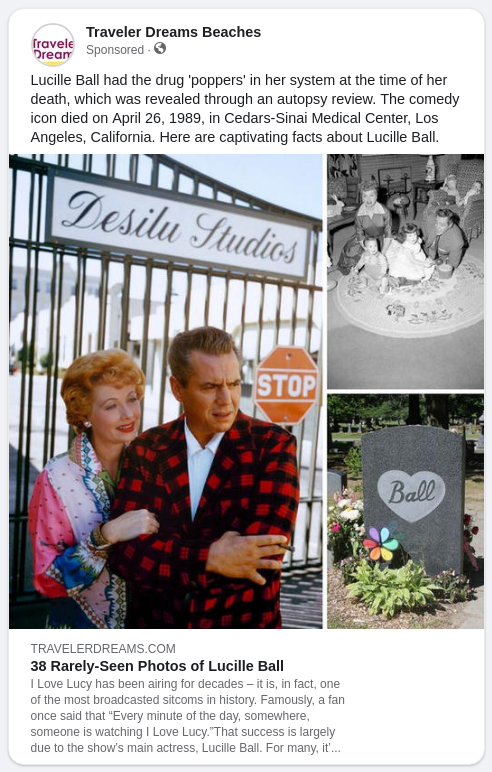}
         \caption{\clickbait{}}
         \label{fig:ad_clickbait}
     \end{subfigure}
     \hfill
     \begin{subfigure}[t]{0.3\textwidth}
         \centering
         \includegraphics[width=\textwidth]{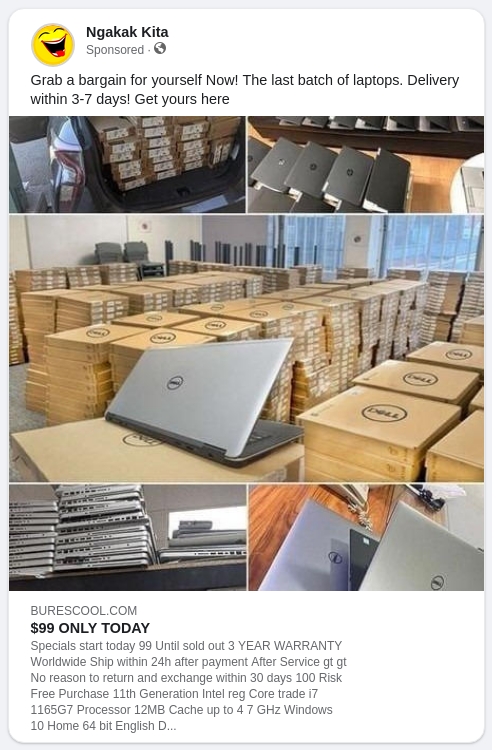}
         \caption{\deceptive{}}
         \label{fig:ad_deceptive}
     \end{subfigure}
     \hfill
     \begin{subfigure}[t]{0.3\textwidth}
         \centering
         \includegraphics[width=\textwidth]{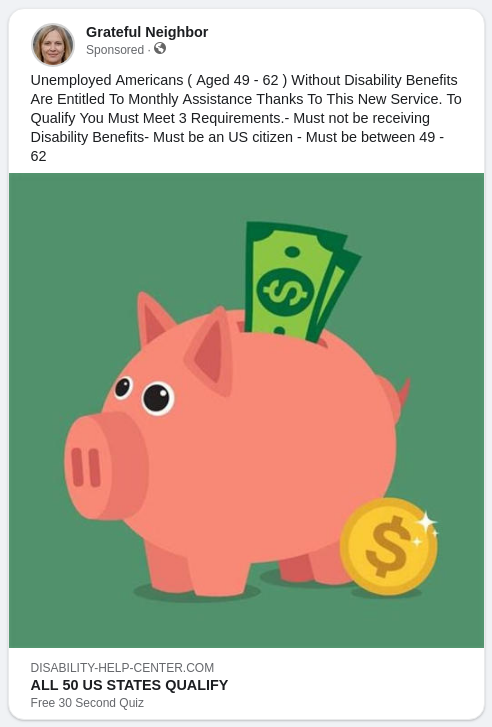}
         \caption{\clickbait{}}
         \label{fig:ad_clickbait}
     \end{subfigure}
     \hfill
     \begin{subfigure}[t]{0.3\textwidth}
         \centering
         \includegraphics[width=\textwidth]{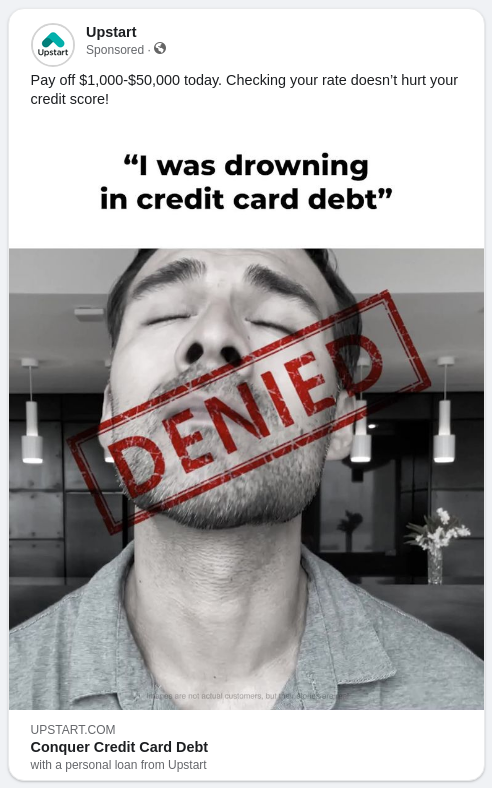}
         \caption{\sensitivefinancial{}}
         \label{fig:ad_financial}
     \end{subfigure}
     \hfill
     \begin{subfigure}[t]{0.3\textwidth}
         \centering
         \includegraphics[width=\textwidth]{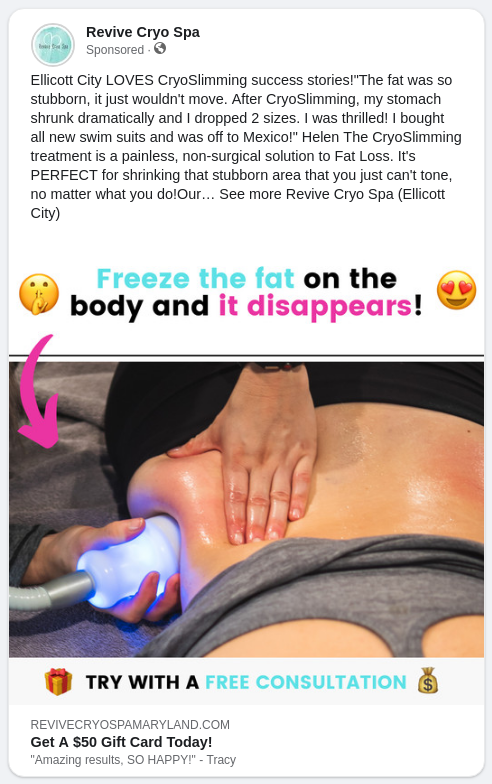}
         \caption{\sensitiveother{}}
         \label{fig:ad_sensitive}
     \end{subfigure}
        \caption{Example images of categories identified as problematic by our participants.}
        \label{fig:ad_images}
\end{figure*}

%% file: tables/joined_table.tex
\begin{table*}[ht]
    % \hspace{50pt}
    \centering
    \renewcommand{\arraystretch}{1}
    \renewcommand\thetable{A1}
    \begin{minipage}{0.64\linewidth}\centering
    \rowcolors{2}{gray!25}{white}
    \resizebox{\textwidth}{!}{
    % \resizebox{\columnwidth}{!}{
    \begin{tabular}{lc|ccccc}
    {\bf Variable} & \multicolumn{6}{c}{\shortstack{\textbf{Estimate ($\beta$)} \\ \scriptsize{[95\% CI]}}} \\
    \toprule
        & {\bf Problematic} & \makecell{\bf Pot. \\\bf Prohibited} & {\bf Deceptive} & {\bf Clickbait} & \makecell{\bf Sensitive: \\\bf Financial} & \makecell{\bf Sensitive: \\\bf Other}\\
        \midrule
    Intercept & \shortstack{0.106$^*$\\ \scriptsize{[0.02, 0.19]}} & \shortstack{0.011$^*$\\ \scriptsize{[0, 0.02]}} & \shortstack{-0.001\\ \scriptsize{[-0.03, 0.03]}} & \shortstack{0.026\\ \scriptsize{[-0.01, 0.06]}} & \shortstack{0.056\\ \scriptsize{[-0.01, 0.13]}} & \shortstack{0.015\\ \scriptsize{[-0.02, 0.05]}} \\ 
    {\bf Gender}: Woman & \shortstack{\textbf{-0.066}$^{***}$\\ \scriptsize{[-0.09, -0.04]}} & \shortstack{-0.002\\ \scriptsize{[0, 0]}} & \shortstack{-0.005\\ \scriptsize{[-0.01, 0]}} & \shortstack{-0.007\\ \scriptsize{[-0.02, 0]}} & \shortstack{\textbf{-0.047}$^{***}$\\ \scriptsize{[-0.07, -0.03]}} & \shortstack{-0.005\\ \scriptsize{[-0.02, 0.01]}} \\ 
    {\bf Race}: Black & \shortstack{0.028$^+$\\ \scriptsize{[0, 0.06]}} & \shortstack{-0.001\\ \scriptsize{[0, 0]}} & \shortstack{0.006\\ \scriptsize{[0, 0.02]}} & \shortstack{\textbf{0.013}$^*$\\ \scriptsize{[0, 0.02]}} & \shortstack{0.007\\ \scriptsize{[-0.02, 0.03]}} & \shortstack{0.003\\ \scriptsize{[-0.01, 0.02]}} \\ 
    {\bf Race}: Asian & \shortstack{-0.001\\ \scriptsize{[-0.04, 0.04]}} & \shortstack{0.001\\ \scriptsize{[0, 0.01]}} & \shortstack{-0.002\\ \scriptsize{[-0.02, 0.01]}} & \shortstack{0.004\\ \scriptsize{[-0.01, 0.02]}} & \shortstack{-0.007\\ \scriptsize{[-0.04, 0.03]}} & \shortstack{0.003\\ \scriptsize{[-0.01, 0.02]}} \\ 
    {\bf Ethnicity}: Hispanic & \shortstack{0.037\\ \scriptsize{[-0.01, 0.09]}} & \shortstack{\textbf{-0.007}$^*$\\ \scriptsize{[-0.01, 0]}} & \shortstack{0.006\\ \scriptsize{[-0.01, 0.02]}} & \shortstack{-0.01\\ \scriptsize{[-0.03, 0.01]}} & \shortstack{\textbf{0.048}$^*$\\ \scriptsize{[0.01, 0.09]}} & \shortstack{0\\ \scriptsize{[-0.02, 0.02]}} \\
    {\bf Education}: college and above & \shortstack{0.009\\ \scriptsize{[-0.02, 0.04]}} & \shortstack{-0.001\\ \scriptsize{[0, 0]}} & \shortstack{0.004\\ \scriptsize{[-0.01, 0.01]}} & \shortstack{\textbf{0.011}$^*$\\ \scriptsize{[0, 0.02]}} & \shortstack{-0.004\\ \scriptsize{[-0.03, 0.02]}} & \shortstack{0\\ \scriptsize{[-0.01, 0.01]}} \\ 
    {\bf Age}: 42 and above & \shortstack{\textbf{0.052}$^{***}$\\ \scriptsize{[0.02, 0.08]}} & \shortstack{\textbf{-0.003}$^*$\\ \scriptsize{[-0.01, 0]}} & \shortstack{\textbf{0.012}$^*$\\ \scriptsize{[0, 0.02]}} & \shortstack{\textbf{0.017}$^{**}$\\ \scriptsize{[0.01, 0.03]}} & \shortstack{0.018\\ \scriptsize{[-0.01, 0.04]}} & \shortstack{0.009\\ \scriptsize{[0, 0.02]}} \\ 
    Privacy Settings Awareness (1--5) & \shortstack{0.003\\ \scriptsize{[-0.02, 0.02]}} & \shortstack{0\\ \scriptsize{[0, 0]}} & \shortstack{0.002\\ \scriptsize{[0, 0.01]}} & \shortstack{-0.003\\ \scriptsize{[-0.01, 0]}} & \shortstack{0.003\\ \scriptsize{[-0.01, 0.02]}} & \shortstack{0.001\\ \scriptsize{[-0.01, 0.01]}} \\   
    \end{tabular}
    }
    %\caption{Coefficients of linear regression models, modeling  similar to Table~\ref{tab:demographic_skew}.\newline $p<0.001^{***}$; $p<0.01^{**}$, $p<0.05^*$, $p<0.1^+$.}
    %\caption{T1}
    % \parbox{.8\textwidth}{
    %     \caption{Coefficients of linear regression models, modeling the relationship between exposure to \problematic{} ads and participants' demographics and privacy behavior, similar to Table~\ref{tab:demographic_skew}.
    %     }
    %     \label{tab:demographic_skew_w_privacy}
    % }
    % \caption{Coefficients of linear regression models, with 95\% confidence intervals, modeling the relationship between exposure to \problematic{} ads and participants' demographics and privacy behavior.
    % Dependent variable (columns): fraction of ad type, out of total ad diet. Independent variable (rows): participant demographics and Likert scale response of their awareness of privacy settings on Facebook.
    % Union of all problematic ad types modeled in the \problematic{} column. \newline $p<0.001^{***}$; $p<0.01^{**}$, $p<0.05^*$, $p<0.1^+$.}
    % \label{tab:demographic_skew_w_privacy}
    \end{minipage}
    \begin{minipage}{0.35\linewidth}\centering
        \rowcolors{2}{gray!25}{white}
        \resizebox{0.95\textwidth}{!}{
            \begin{tabular}{lcc}
            {\bf Variable} & \multicolumn{2}{c}{\shortstack{\textbf{Estimate ($\beta$)} \\ \scriptsize{[95\% CI]}}} \\
            \toprule
            & {\bf Healthcare} & {\bf Opportunity} \\
            \midrule
             Intercept & \shortstack{0.089$^{***}$\\ \scriptsize{[0.06, 0.11]}} & \shortstack{0.045$^{**}$ \vspace{.2em}\\ \scriptsize{[0.01, 0.08]}} \\ 
              {\bf Gender}: Woman & \shortstack{0.007\\ \scriptsize{[-0.01, 0.03]}} & \shortstack{0.022$^+$\\ \scriptsize{[0, 0.05]}} \vspace{.2em}\\ 
              {\bf Race}: Black & \shortstack{-0.024$^+$\\ \scriptsize{[-0.05, 0]}} & \shortstack{\textbf{0.038}$^*$\\ \scriptsize{[0.01, 0.07]}} \vspace{.2em}\\ 
              {\bf Race}: Asian & \shortstack{-0.017\\ \scriptsize{[-0.05, 0.01]}} & \shortstack{0.028\\ \scriptsize{[-0.01, 0.07]}} \vspace{.2em}\\ 
              {\bf Education}: college and above& \shortstack{-0.004\\ \scriptsize{[-0.03, 0.02]}} & \shortstack{\textbf{0.034}$^*$\\ \scriptsize{[0.01, 0.06]}} \vspace{.2em}\\ 
              {\bf Ethnicity}: Hispanic & \shortstack{0.011\\ \scriptsize{[-0.03, 0.05]}} & \shortstack{-0.007\\ \scriptsize{[-0.06, 0.04]}} \vspace{.2em}\\ 
              {\bf Age}: 42 and above & \shortstack{0.018$^+$\\ \scriptsize{[0, 0.04]}} & \shortstack{-0.021\\ \scriptsize{[-0.05, 0.01]}} \vspace{.2em}\\   
               %\midrule
               %$R^2$ & 0.088 & 0.151 \\
        \end{tabular}
        }
        %\caption{T2}
        %\caption{Coefficients of two regression models, modeling the relationship between exposure to Healthcare and Opportunity ads, and participant demographics. Analysis setup similar to Table~\ref{tab:demographic_skew}. \newline $p<0.001^{***}$; $p<0.01^{**}$, $p<0.05^*$, $p<0.1^+$.}
    \end{minipage}
    \caption{Coefficients of linear regression models. Left: modeling the relationship between exposure to \problematic{} ads and participants' demographics and privacy behavior.
    Right: modeling the relationship between exposure to \healthcare{} and \opportunity{} ads, and participant demographics. \newline $p<0.001^{***}$; $p<0.01^{**}$, $p<0.05^*$, $p<0.1^+$.}
    \label{tab:demographic_skews_extra}
\end{table*}

%% file: tables/interest_targeting_condensed.tex
\begin{table*}[b]
\centering
\rowcolors{2}{gray!25}{white}
\small
{\def\arraystretch{1.2}\tabcolsep=10pt
\renewcommand\thetable{A2}
\begin{tabular}{rp{13.5cm}}
\textbf{Ad Category} & \textbf{Targeted Interest (Prevalence)} \\ \toprule
Neutral & {\small \textbf{None} (\textbf{72.3\%}), Online shopping (1.3\%), Health \& wellness (0.8\%), Family (0.7\%), Physical fitness (0.7\%), Yoga (0.6\%)}\\
Opportunity & {\small \textbf{None} (\textbf{67.9\%}), Employment (2.7\%), Education (2.4\%), Higher education (2.3\%), Career (1.7\%), Technology (1.6\%)}\\
Healthcare & {\small \textbf{None} (\textbf{76.8\%}), Health \& wellness (2.4\%), Clinical trial (2.2\%), Physical fitness (1.7\%), Physical exercise (1.5\%), Medicine (1.0\%)}\\
Clickbait & {\small \textbf{None} (\textbf{79.8\%}), Online shopping (1.3\%), Personal finance (1.0\%), Amazon.com (0.9\%), Home improvement (0.8\%), Investment (0.8\%)}\\
Sensitive: Financial & 	{\small \textbf{None} (\textbf{76.4\%}), Personal finance (5.1\%), Investment (3.4\%), Online banking (3.1\%), Credit cards (2.9\%), Financial services (2.0\%)}\\
Sensitive: Other & {\small \textbf{None} (\textbf{75.3\%}), Gambling (2.9\%), Alcoholic beverages (2.4\%), Bars (2.1\%), Beer (1.9\%), Vodka (1.4\%)}\\
Pot. Prohibited & {\small \textbf{None} (\textbf{81.3\%}), Health \& wellness (2.7\%), Meditation (1.8\%), Physical fitness (1.4\%), Credit cards (1.4\%), House Hunting (1.4\%)}\\
Deceptive  & {\small \textbf{None} (\textbf{68.1\%}), Online shopping (4.0\%), Shopping (2.1\%), Amazon.com (1.5\%), Clothing (1.5\%), Digital marketing (1.5\%)}\\
\end{tabular}
% }
\caption{Most popular targeting interests by category. We see that a majority of ads are not targeted by interests.}
\label{tab:targeting_interests}
}
\end{table*}

%% file: arxiv-ae.tex
%%%%%%%%%%%%%%%%%%%%%%%%%%%%%%%%%%%%%%%%%%%%%%%%%%%%
% Artifact Appendix Template for EuroSys'22 AE
%
% this document has a maximum length of 2 pages.
%%%%%%%%%%%%%%%%%%%%%%%%%%%%%%%%%%%%%%%%%%%%%%%%%%%%

% \appendix
% \section{Artifact Appendix}
\clearpage
\setcounter{figure}{0}
\renewcommand\thefigure{B\arabic{figure}} 
\renewcommand\thesection{B}
\section{Codebook}

%%%%%%%%%%%%%%%%%%%%%%%%%%%%%%%%%%%%%%%%%%%%%%%%%%%%%%%%%%%%%%%%%%%%%
% \subsection{Abstract}
This appendix replicates the codebook used by annotators to manually categorize ads.
Each code was described through a definition, some descriptive examples, and some examples of ads that would not qualify as the code. An example ad is also provided for each code.

%%%%%%%%%%%%%%%%%%%%%%%%%%%%%%%%%%%%%%%%%%%%%%%%%%%%%%%%%%%%%%%%%%%%%
\subsection{Deceptive}

\parc{Definition} Ads that may overtly or deceptively lead users to engage with fraudulent offers, potential scams, false or misleading claims, or predatory business practices (e.g. recurring billing). 

\begin{figure}[h]
  \centering
  \includegraphics[width=.6\linewidth]{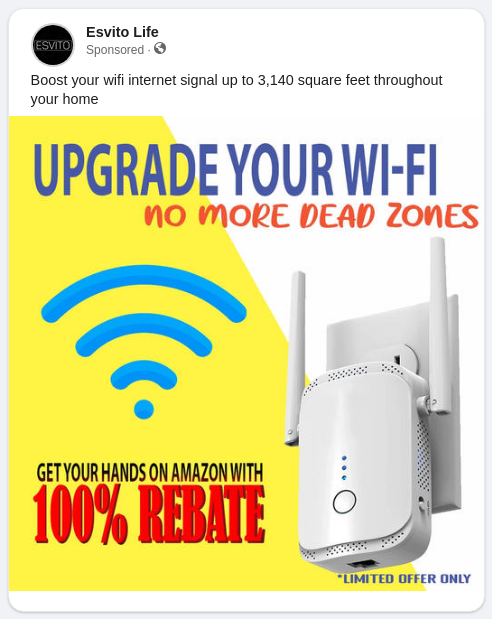}
\caption{Example \deceptive{} ad; multiple reviews on Facebook page mention a rebate was never issued.}
\label{fig:deceptive}
\end{figure}

\parc{Qualifies}
\begin{packed_itemize}
    \item Ads that have potential to harm users financially, such as
    \begin{packed_itemize}
        \item Payday loans, paycheck advances, bail bonds, or any short-term loans\footnote{Also considered deceptive in Facebook's policies: \url{https://transparency.fb.com/policies/ad-standards/deceptive-content/}}
        \item Debt settlement services, especially with strongly worded guarantees
    \end{packed_itemize} 
    \item Services which are highly unlikely to result in the advertised outcome, e.g. fat burning pills, guaranteed monthly income etc.
    \item Scams---either for money or personal information---that we can confirm via Facebook reviews or Better Business Bureau reports
    \item Ads that employ deceptive tactics, such as:    
    \begin{packed_itemize}
        \item Containing false or exaggerated claims
        \item Designed to look like they are advertising a different product than the linked webpage
    \end{packed_itemize}
    \item Overly pushy or manipulative ads
    \item Predatory business practices such as: requests for direct messages, recurring/non-cancellable billing as mentioned by users
\end{packed_itemize}

\parc{Does not qualify}
\begin{packed_itemize}
    \item Sketchy product ads for which we cannot find evidence of deception
    \item Visibly low quality products that are not inherently harmful e.g. clothing, jewellery, novels
\end{packed_itemize}

\subsection{Potentially Prohibited}

Ads that might be prohibited according to Facebook’s prohibited policies\footnote{\url{https://transparency.fb.com/policies/ad-standards/}} on unacceptable content, dangerous content and objectionable content.

\begin{figure}[h]
  \centering
  \includegraphics[width=.6\linewidth]{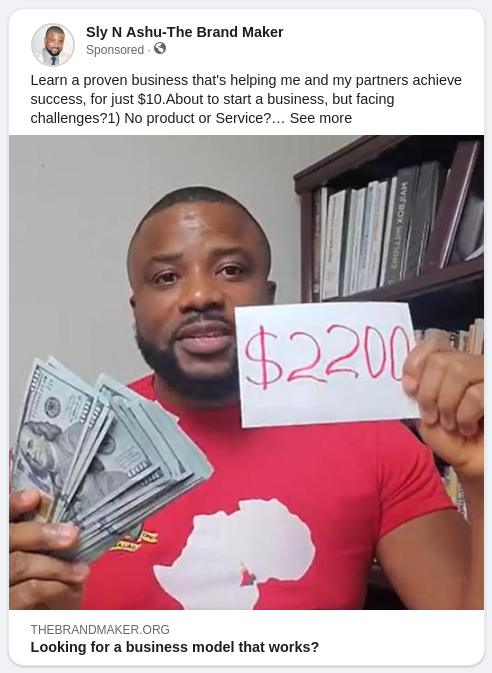}
\caption{Example \prohibited{} ad; advertises a business plan that offers quick income with low investment.}
\label{fig:prohibited}
\end{figure}

\parc{Qualifies}
\begin{packed_itemize}
    \item Prohibited substances---such as illegal prescription and recreational drugs, tobacco, and related products
    \item Unsafe dietary supplements and medical treatments
    \item Weapons, ammunition or explosives
    \item Adult products or services
    \item Instant loans, pre-financing and security deposits
    \item Selling human body parts or fluids
    \item Multilevel marketing, or income opportunities that offer quick income with low investment
    \item Spyware or malware
    \item Ads against vaccinations   
\end{packed_itemize}

\parc{Does not qualify}
\begin{packed_itemize}
    \item Products for sexual or reproductive health, such as medical devices for family planning and contraception
\end{packed_itemize}

\subsection{Clickbait}
\parc{Definition} Clickbait intentionally omits crucial information or exaggerates the details of a story to make it seem like a bigger deal than it really is\footnote{Facebook Business Center: \url{https://www.facebook.com/business/help/503640323442584?id=208060977200861}}. Such ads often have three distinct characteristics~\cite{zeng2021makes}: the ad is attention grabbing, the ad does not tell the viewer exactly what is being promoted to ``bait'' the viewer into clicking it, and the landing page of the ad often does not live up to people’s expectations based on the ad.

\begin{figure}[h!]
  \centering
  \includegraphics[width=.6\linewidth]{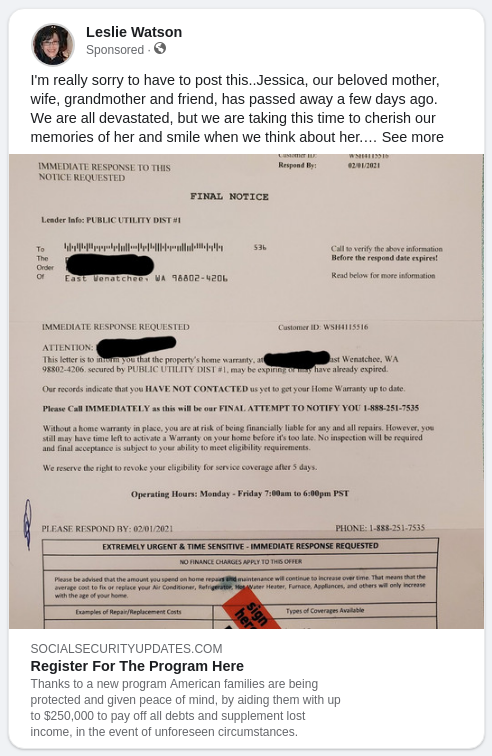}
\caption{Example \clickbait{} ad.}
\label{fig:clickbait}
\end{figure}

\parc{Qualifies}
\begin{packed_itemize}
    \item Ads where the text, headline, and description together don’t clarify what precisely is being advertised
    \item Ads that omit information to entice users
    \item Have very dense text in the image
    \item Invite the users to tap a section of the image/button in image for results
    \item Ads for products that are actually affiliate marketing links or data collection forms, e.g. home renovation and solar panel surveys
\end{packed_itemize}

\parc{Does not qualify}
\begin{packed_itemize}
    \item Ads that use loud language but are clear about what the advertised product, e.g. e-commerce ads that advertise 200\% growth in business but are upfront that they’re advertising an online course
\end{packed_itemize}

\subsection{Sensitive: Financial}
\parc{Definition} Ads that contain products or services related to managing finances, building credit, and other financial tools. Such ads are subject to content-specific restrictions on Facebook\footnote{\url{https://transparency.fb.com/policies/ad-standards/content-specific-restrictions}}, and must comply with additional targeting and authorization restrictions.

\begin{figure}[h!]
  \centering
  \includegraphics[width=.6\linewidth]{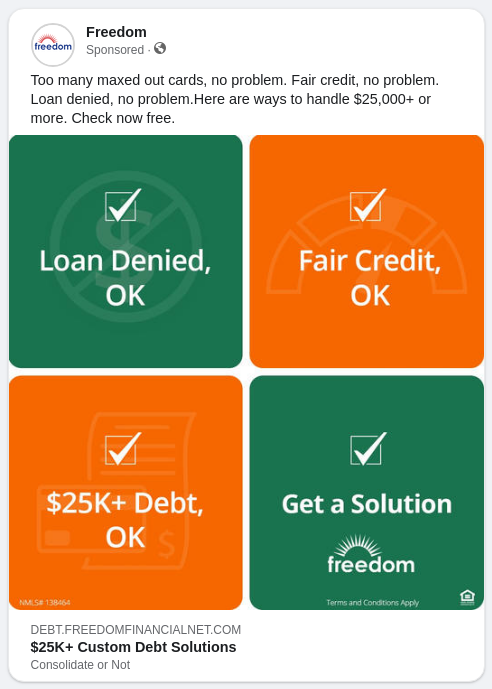}
\caption{Example \sensitivefinancial{} ad.}
\label{fig:financial}
\end{figure}

\parc{Qualifies}
\begin{packed_itemize}
    \item Loans and mortgage financing ads
    \item Checking, savings and brokerage accounts ads
    \item Credit card ads
    \item Ads regarding financial and investment advice
    \item Cryptocurrency or stock investment opportunities
\end{packed_itemize}

\parc{Does not qualify}
\begin{packed_itemize}
    \item Ads that reference saving money, but whose products or services are not inherently financial (e.g. browser extensions for deal-hunting)
    \item Insurance advertising
    \item Ads for selling homes
\end{packed_itemize}

\subsection{Sensitive: Other}
\parc{Definition} Ads that contain subject matter that may be sensitive or triggering for users to view, or they may contain content that is harmful for vulnerable groups (e.g. those suffering from an addiction). Similar to \sensitivefinancial{}, such ads are subject to content-specific restrictions on Facebook, and must comply with additional targeting and authorization restrictions.

\begin{figure}[h!]
  \centering
  \includegraphics[width=.6\linewidth]{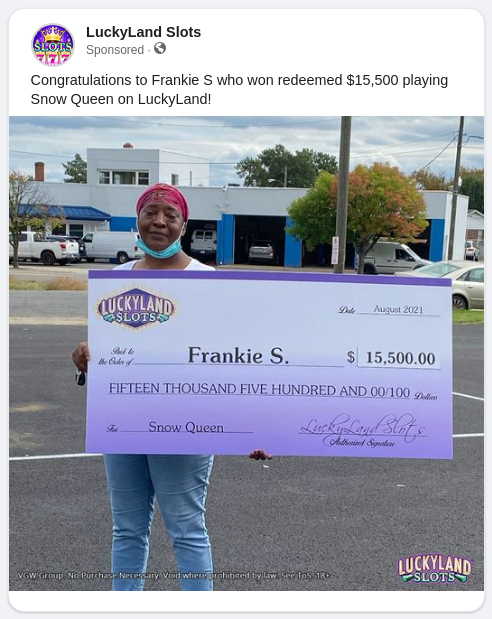}
\caption{Example \sensitiveother{} ad.}
\label{fig:sensitive}
\end{figure}

\parc{Qualifies}
\begin{packed_itemize}
    \item Alcohol
    \item Gambling, casinos, online slot machines
    \item Dieting, weight loss treatments, anything related to body image
    \item Prescription and over-the-counter drugs
    \item Online pharmacies and services for mental and physical health
\end{packed_itemize}

\parc{Does not qualify}
\begin{packed_itemize}
    \item Ads for meal plans
    \item Popular brick and mortar pharmacy ads e.g., CVS and Walgreens
\end{packed_itemize}

\subsection{Opportunity}
\parc{Definition} Ads that present any sort of employment, housing, or educational opportunity to users.

\begin{figure}[h!]
  \centering
  \includegraphics[width=.6\linewidth]{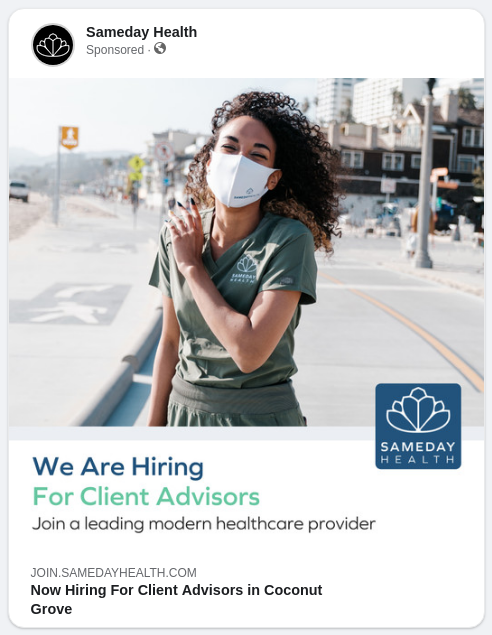}
\caption{Example \opportunity{} ad; advertises a job.}
\label{fig:opportunity}
\end{figure}

\parc{Qualifies}
\begin{packed_itemize}
\item Ads for a job or gig, such as
\begin{packed_itemize}
    \item Full- or part-time employment opportunities
    \item Gig-work opportunities like Uber, Doordash etc.
    \item Local job fairs
\end{packed_itemize}
\item Ads for educational opportunities (for-profit universities and online degrees included)
\item Fellowships, scholarships, writing contests, etc.
\end{packed_itemize}

\parc{Does not qualify}
\begin{packed_itemize}
\item Product sweepstakes or cash-back promotions
\item Ads with financial opportunities (savings, credit cards etc.), regardless of how big the sign-up rewards are
\item Online studies or market research opportunities, regardless of compensation
\end{packed_itemize}

\subsection{Healthcare}
\parc{Definition} Ads that contain products or services related to healthcare, fitness, mental and physical wellness, or physical appearance. 

\begin{figure}[h!]
  \centering
  \includegraphics[width=.6\linewidth]{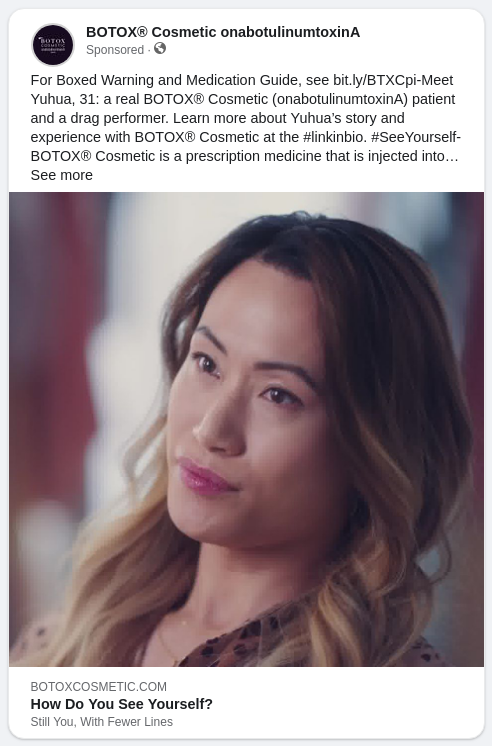}
\caption{Example \healthcare{} ad; advertises a cosmetic treatment.}
\label{fig:healthcare}
\end{figure}

\parc{Qualifies}
\begin{packed_itemize}
    \item Fitness products or services, and gym memberships
    \item Vitamins or supplements
    \item Physical appearance-related products or services, like hair growth supplements
    \item Health insurance
    \item Dieting products or services (also annotated as \sensitive{})
    \item Online mental health clinics and prescription services
    \item At-home medical monitoring devices
    \item Public-health announcements (e.g. CDC, WHO)
    \item Ads for children’s health
\end{packed_itemize}

\parc{Does not qualify}
\begin{packed_itemize}
    \item Ads for meal plans
    \item Ads for pet health
\end{packed_itemize}

\subsection{Political}
\parc{Definition} Ads containing references to political parties, bills or laws, political figures or candidates, petitions or causes with outwardly political affiliations, or any other political-related content.

\begin{figure}[h!]
  \centering
  \includegraphics[width=.6\linewidth]{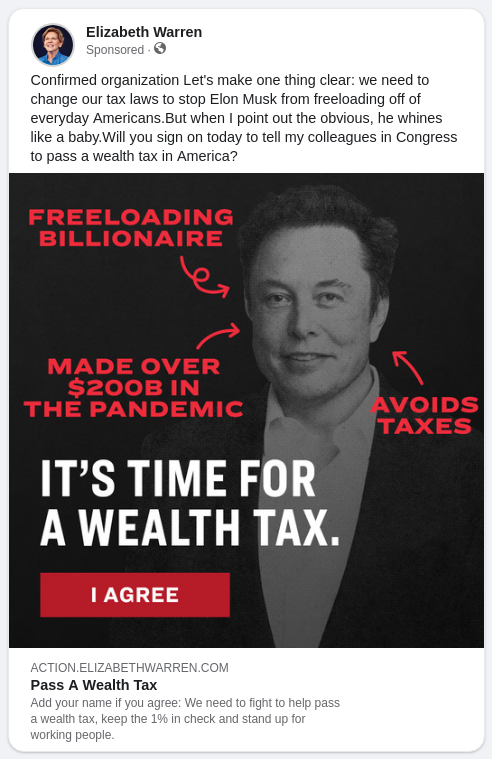}
\caption{Example \political{} ad.}
\label{fig:political}
\end{figure}

\parc{Qualifies}
\begin{packed_itemize}
    \item References to any political candidates or figures
    \item References to any bills, laws, legislation, etc.
    \item Petitions, causes, events, or fundraisers that are politically affiliated or motivated
    \item References to political parties
\end{packed_itemize}

\parc{Does not qualify}
\begin{packed_itemize}
    \item Ads for political merchandise, e.g. t-shirts
    \item Religious ads    
\end{packed_itemize}

\subsection{Neutral}
\parc{Definition} Ads that simply seek to advertise a product, service, local event, apolitical news etc. Ads that don’t fall into other categories, and seem benign based on their impact on users, should be marked \neutral. This code is mutually exclusive from all others.

\begin{figure}[h!]
  \centering
  \includegraphics[width=.6\linewidth]{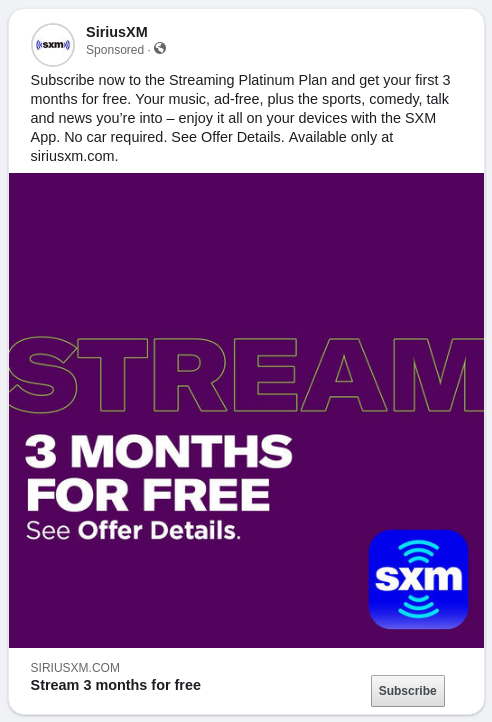}
\caption{Example \neutral{} ad.}
\label{fig:neutral}
\end{figure}

\parc{Qualifies}
\begin{packed_itemize}
    \item Product ads
    \item Services and subscriptions
    \item News and information ads, from news outlets as well
    \item Religious ads
    \item Insurance ads
\end{packed_itemize}

\parc{Does not qualify}
\begin{packed_itemize}
    \item Ads that fit strongly into one of our other specific categories
    \item Ads with either opportunities or potential harmful outcomes for users    
\end{packed_itemize}

% \subsection{Version}
% %%%%%%%%%%%%%%%%%%%%
% % Obligatory.
% % Do not change/remove.
% %%%%%%%%%%%%%%%%%%%%
% Based on the LaTeX template for Artifact Evaluation V20220926. Submission,
% reviewing and badging methodology followed for the evaluation of this artifact
% can be found at \url{https://secartifacts.github.io/usenixsec2023/}.